\documentclass[aps,prb,twocolumn,showpacs,floatfix,twoside,superscriptaddress]{revtex4-2}
\usepackage{amssymb}
\usepackage{graphicx}
\usepackage{amsmath}
\usepackage{amsfonts}
\usepackage{bbold}
\usepackage{esint}
\usepackage{color}
\usepackage[dvipsnames]{xcolor}
\usepackage{float}
\usepackage{verbatim}
\usepackage[colorlinks=true,citecolor=blue,linkcolor=blue,urlcolor=blue]{hyperref}
\usepackage[justification=centerlast,font={small}]{caption}

\DeclareMathOperator*{\tr}{Tr}
\DeclareMathOperator{\im}{Im}
\DeclareMathOperator{\re}{Re}

\newcommand{\BT}{\mathcal{T}}

\newcommand{\BP}{\mathcal{P}}

\begin{document}
 
 \title{Schrieffer-Wolff transformation for non-Hermitian systems: application for $\BP\BT$--symmetric circuit QED
 }
\author{Grigory A. Starkov}
\email{Grigorii.Starkov@rub.de}
\author{Mikhail V. Fistul}
\author{Ilya M. Eremin}
\affiliation{Institut fur Theoretische Physik III, Ruhr-Universitat Bochum, 44801 Bochum, Germany}

\date{\today}

\begin{abstract}
 Combining non-hermiticity and interactions yields novel effects in open quantum many-body systems. Here, we develop the generalized Schrieffer-Wolff transformation and derive  the effective Hamiltonian suitable for various quasi-degenerate \textit{non-Hermitian} systems. We apply our results to an exemplary  $\BP\BT$--symmetric circuit QED composed of two 
non-Hermitian qubits embedded in a lossless resonator. We consider a resonant quantum circuit as $|\omega_r-\Omega| \ll \omega_r$, where $\Omega$ and $\omega_r$ are qubits and resonator frequencies, respectively, providing well-defined groups of quasi-degenerate resonant states.  For such a system, using direct numerical diagonalization we obtain the dependence of the low-lying eigenspectrum on the interaction strength between a single qubit and the resonator, $g$, and the gain (loss) parameter $\gamma$, and compare that with the eigenvalues obtained analytically using the effective Hamiltonian of resonant states. We identify $\BP\BT$--symmetry broken and unbroken phases, trace the formation of Exceptional Points of the second and the third order, and provide a complete phase diagram $g-\gamma$ of low-lying resonant states. 
We relate the formation of Exceptional Points to the additional $\BP$-pseudo-Hermitian symmetry of the system and show that non-hermiticity mixes the "dark" and the "bright" states, which has a direct experimental consequence.
\end{abstract}

\maketitle

\section{Introduction}
Seminal Schrieffer-Wolff transformation \cite{FoldyWouthhuysen-1950,SchriefferWolff-1966,Bravyi-2011}(also known as Van Vleck Quasi-Degenerate Perturbation theory~\cite{VanVleck-1929,Jordahl-1934,Kemble-1937,Brandow-1979,Hoffmann-1996} in some other contexts)
has been used for many years as an extremely useful tool in a theoretical study of weakly interacting quantum systems. The Schrieffer-Wolff transformation allows one to consistently derive the effective low-energy Hamiltonian composed of the non-interacting 
part and the interaction terms taken into account as a perturbation.
Generally, Schrieffer-Wolf transformation was successfully applied to the perturbation analysis of various quantum-mechanical problems arising in  electronic or optical systems, such as the Kondo problem in the Anderson impurity model in normal metals or superconductors \cite{SchriefferWolff-1966,salomaa1988schrieffer}, multiorbital Hubbard models \cite{lee2017generalized}, Floquet spectrum of periodically driven systems \cite{bukov2016schrieffer}, interacting qubits \cite{blais2004cavity,krantz2019quantum,zagoskin2013spatially,zhang2022quantum,consani2020effective,roth2019adiabatic,blais2021circuit}. 
Moreover, the Schrieffer-Wolff transformation was also adapted to intrinsically dissipative quantum systems described by Markovian master equations \cite{kessler2012generalized} and, more recently, discussed in the context of some non-hermitian systems \cite{Lourenco218,Masarelli2022}.

A particular example of the qubits-type systems coined as \textit{circuit QED}, i.e., various arrays of superconducting qubits coupled to a low dissipative resonator,  presents a special interest for quantum computing, quantum simulations, and precise quantum measurements \cite{blais2004cavity,krantz2019quantum,blais2021circuit,acin2018quantum}. Even a weak coupling of qubits to the resonator leads to numerous interesting phenomena, e.g., AC Stark shift of qubits frequencies \cite{blais2004cavity,wallraff2004strong}, or the interaction between well separated qubits mediated by the exchange of virtual photons of the resonator \cite{blais2004cavity,van2018microwave,fink2009dressed},  and the Schrieffer-Wolff transformation are routinely used to analyze these effects. 

Recently, this field of study has got a new twist as so-called $\BP\BT$--symmetric non-Hermitian qubits systems have been experimentally realized in various solid state systems such as trapped ions, ultracold and Rydberg atoms \cite{ding2021experimental,lourencco2022non,li2019observation}, Bose–Einstein condensate \cite{cartarius2012model}, superconducting \cite{naghiloo2019,chen2021quantum,dogra2021quantum} or nitrogen-vacancies qubits \cite{wu2019observation}.
In a $\BP\BT$--symmetric non-Hermitian qubit,  specially chosen
quantum states demonstrate a nonequilibrium growth of the population, i.e., the states with \textit{a gain}. To establish the $\BP\BT$--symmetry the gain has to be completely equalized by a loss present in other parts of a system. 

A general theoretical study of $\BP\BT$-symmetric non-Hermitian Hamiltonians has been started by the works of Bender and coworkers \cite{bender1998real,bender1999pt,bender2007making}. They have shown that depending on the values of physical parameters the $\BP\BT$--symmetric non-Hermitian Hamiltonian exhibits two kinds of energy spectrum: purely real eigenvalues, identifying the unbroken (preserved) $\BP\BT$-symmetric quantum phase, or complex conjugate ones, indicating the appearance of broken $\BP\BT$--symmetry quantum phase. The Exceptional Points (\textit{EP}s) separate these phases. 

Systems of $\BP\BT$-symmetric interacting  qubits (two-levels or spin $1/2$) 
have been theoretically studied for two $\BP\BT$-- non-Hermitian qubits with an exchange interaction \cite{tetling2022linear} as well as in the framework of the quantum Ising model with a longitudinal \cite{starkov2023quantum,PhysRevA.108.022206} or transverse non-Hermitian fields \cite{Song-15}. At the same time, one should stress that in most previous works a theoretical study of the behavior of eigenvalues and eigenvectors near an $n$-th order \textit{EP} has been carried out using the \textit{local} effective $n \times n$ Hamiltonian. Therefore, a study of the relation between the \textit{EP}s arising in the presence of interaction between qubits, and the properties of non-interacting qubits system did not receive a proper attention so far. 

In this Article, we develop the generalized Schrieffer-Wolff transformation for arbitrary non-Hermitian interacting systems and derive the effective Hamiltonian in which the interaction is treated as the perturbation up to the second order. Since the Schrieffer-Wolff transformation is especially suited for quantum systems with well-defined groups of quasi-degenerate states \cite{Bravyi-2011}, we apply our generic analysis to an exemplary $\BP\BT$--symmetric quantum system, i.e., a $\BP\BT$--symmetric circuit QED composed of two non-Hermitian qubits weakly coupled to a lossless resonator. The non-Hermiticity is introduced in the circuit QED model in the form of an imaginary longitudinal magnetic field of a strength, $\gamma$, or equivalently as a gain (loss) of corresponding energy levels of a single qubit. The resonant regime will be addressed as the qubit's frequency $\Omega$ is assumed to be close to the frequency of a resonator, $\omega_r$ \cite{blais2004cavity,blais2021circuit}. Combining the direct numerical diagonalization with the analysis based on the derived effective Hamiltonian, we study the low-energy spectrum of $\BP\BT$--symmetric circuit QED in detail. We identify various $\BP\BT$-symmetry preserved and broken quantum phases, and quantitatively analyze the formation of \textit{EP}s of different orders. 

The outline of the paper is as follows.
In Section~\ref{section:transformation} we formulate  the general Schrieffer-Wolff transformation suitable for quasi-degenerate non-Hermitian Hamiltonian systems. Using such transformation, we derive the effective Hamiltonian. 
In Section~\ref{section:model} we introduce a basic model of $\BP\BT$--symmetric circuit QED, i.e., two non-Hermitian qubits embedded in a lossless resonator, discuss important symmetries of a system, and, using the direct numerical diagonalization, obtain the 
dependence of the eigenspectrum on the coupling strength, $g$, and the gain (loss) parameter, $\gamma$. In Section~\ref{section:effham}, we use the Schrieffer-Wolff transformation to derive the effective Hamiltonian matrix of the resonant states and discuss its symmetry properties. After that, in Section~\ref{section:effham-analysis}, using the effective Hamiltonian we study the formation of Exceptional Points in detail, and produce a complete phase diagram of the low-lying resonant states. In Section~\ref{section:conclusions}  the conclusions will be provided. 
There are also Appendix~\ref{section:matelems}, where we compute the matrix elements of 
the effective Hamiltonian, and Appendices ~\ref{section:ch-pol-sol}, \ref{section:critvals} and ~\ref{section:eigvals-perturb}, where the derivation details regarding the phase diagram and \textit{EP}s are presented. 

\section{Schrieffer-Wolff transformation for a quasi-degenerate non-Hermitian Hamiltonian\label{section:transformation}}

Let us consider a quantum system described by an arbitrary non-Hermitian Hamiltonian of the form
\begin{equation}
 \hat H = \hat H_0 + g\hat V,\label{ham-perturb}
\end{equation}
where both $\hat H_0$ and $\hat V$ can be non-Hermitian. 
We assume that the main part of the Hamiltonian, $\hat H_0$, has a well-defined group of quasi-degenerate eigenstates 
which is clearly separated from other levels in energy. The parameter $g$ is considered to be small, and we will treat the second term $g\hat V$ as a perturbation.

We further assume, that $\hat H_0$ is not close to the \textit{EP}s and it has a complete biorthonormal basis of right $|R_i\rangle$ and left $\langle L_i|$ eigenvectors~\cite{mostafazadeh2010}:
\begin{equation}
  \hat H_0 |R_i\rangle = E_i^{(0)} |R_i\rangle,\qquad \langle L_i|\hat H_0 = E_i^{(0)} \langle L_i|,
\end{equation}
\begin{equation}
 \langle L_i| R_j\rangle = \delta_{i,j}.
\end{equation}
These eigenvectors satisfy the condition ~\cite{mostafazadeh-10}
\begin{equation}
 \mathbb{1} = \sum_i |R_i\rangle\langle L_i|\label{id-resol}.
\end{equation}
Let $P$ be the set of quasi-degenerate eigenstates of the Hamiltonian $\hat H_0$ that we want to project onto, and let $Q = \bar P$ be the set of all other states.
In the non-Hermitian case, the projector operator $\hat P$ on the set of  states $P$ is defined as~\cite{Ashida-20,kessler2012generalized}:
\begin{equation}
 \hat P = \sum_{p\in P} |R_p\rangle\langle L_p|.\label{pproj}
\end{equation}
We also introduce the orthogonal projector:
\begin{equation}
 \hat Q = \mathbb{1} - \hat P\label{orthprojdef}
\end{equation}
written explicitly as 
\begin{equation}
\hat Q = \sum_{q\in Q} |R_q\rangle\langle L_q|.\label{qproj}
\end{equation}
Using the
biorthogonality of the left and right eigenvectors, one can check that the projector operators $\hat P$ and $\hat Q$ satisfy the following properties: $\hat P^2=\hat P$, $\hat Q^2=\hat Q$ and $\hat P \hat Q = \hat Q\hat P = 0$.

The use of the projector operators, $\hat P$ and $\hat Q$, allows one to split the Hamiltonian as~\cite{Brandow-1979, Hoffmann-1996}:
\begin{equation}
 \hat H = \hat H_0 + g\underbrace{\left[\hat P \hat V \hat P + \hat Q \hat V \hat Q\right]}_{\hat V_D} + g\underbrace{\left[\hat P \hat V \hat Q + \hat Q \hat V \hat P\right]}_{\hat V_X}.\label{proj-split}
\end{equation}
Note, the projector operators $\hat P$ and $\hat Q$ 
commute with $\hat H_0$. As a consequence, $\hat P\hat H_0\hat Q = \hat Q\hat H_0\hat P=0$, and
\begin{equation}
 \hat H_0 = \hat P \hat H_0 \hat P + \hat Q \hat H_0 \hat Q\label{zero-split}
\end{equation}
The operator $\hat H_0 + g \hat V_D$ yields the "diagonal" part of the Hamiltonian that does not mix the states from subspaces $P$ and $Q$.  On the contrary, the operator $g\hat V_X$ gives rise to the "off-diagonal" part of the Hamiltonian connecting the states from different subspaces.

Next, we search for the transformation allowing to get rid of  
the off-diagonal part $g\hat V_X$ of the Hamiltonian~\eqref{proj-split}. 
More precisely, one have to find the operator $\hat S$, such that
\begin{equation}
 e^{g\hat{S}} \left[\hat H_0 + g\hat V\right] e^{-g\hat S} = \hat H^\prime,\label{transform}
\end{equation}
where $\hat H^\prime$ satisfies
\begin{equation} \label{transform-1}
    \hat P \hat H^\prime \hat Q = \hat Q\hat H^\prime \hat P = 0.
\end{equation}
The procedure described by Eqs. (\ref{transform}) and (\ref{transform-1}), defines the Schrieffer-Wolff transformation. 
For Hermitian Hamiltonian $\hat H$, the operator $\hat S$ is an anti-Hermitian one, and Eq.~\eqref{transform} describes a unitary transformation. For a non-Hermitian case, the operator $\hat S$ acquires a Hermitian component, and Eq.~\eqref{transform} determines then a similarity transformation.

As usual, Eq.~\eqref{transform} is written as the series of chained commutators:
\begin{equation}
 e^{g\hat S} \hat H e^{-g\hat S} = \hat H + \sum_{n=1}^{+\infty} \frac{g^n}{n!} \underbrace{\left[\hat S, \left[\hat S,\dotsc\left[\hat S\right.\right.\right.}_{n\ \mathrm{times}}, \left.\left.\left.\hat H\right]\right]\right].\label{commutator-expansion}
\end{equation}
Expanding the operator $\hat S$ in a series of powers of $g$ as
\begin{equation}
 \hat S = \hat S_0 + g\hat S_1 + g^2\hat S_2 +\dotsc,
\end{equation}
and substituting it in  Eq.~\eqref{commutator-expansion} we express the transformed Hamiltonian $\hat H^\prime$ also as  a series in powers of $g$.
Substituting the Hamiltonian~\eqref{proj-split} into Eq.~\eqref{commutator-expansion} and keeping only the terms up to the second order in $g$, we find the \textit{effective } Hamiltonian $\hat H_\mathrm{eff}$ in the following form
\begin{multline}
 \hat H_\mathrm{eff} = \hat H_0 +g\left\{\hat V_D + \hat V_X + \left[\hat S_0,\hat H_0\right]\right\}+ \\ +
 g^2\left\{\left[\hat S_1, \hat H_0\right] + \left[\hat S_0, \hat V_D\right]
 +\vphantom{\frac{1}{2}}\right.\\\left.+
 \left[\hat S_0,\hat V_X\right]+\frac{1}{2}\left[\hat S_0,\left[\hat S_0,\hat H_0\right]\right]\right\}.\label{effective-ham-proj}
\end{multline}

To eliminate $\hat V_X$ in the first order, we require
\begin{equation}
 \left[\hat S_0,\hat H_0\right]=-\hat V_X.
\end{equation}
and $\hat S_0$ is obtained  explicitly as 
\begin{equation} \label{SO-operator}
 \hat S_0 = \hat P \hat U_1 \hat Q + \hat Q\hat U_2\hat P,
\end{equation}
where
\begin{equation}
 \hat U_1 = -\sum_{p\in P, q\in Q} \frac{|R_p\rangle\langle L_p|\hat V|R_q\rangle\langle L_q|}{E_q^0- E_p^0},
\end{equation}
\begin{equation}
 \hat U_2 = \sum_{q\in Q, p\in P} \frac{|R_q\rangle\langle L_q|\hat V|R_p\rangle\langle L_p|}{E_q^0- E_p^0}.
\end{equation}
Choosing the operator $\hat S_1$ to satisfy the condition $[\hat S_1, \hat H_0] + [\hat S_0, \hat V_D]=0$ we obtain the effective Hamiltonian as
\begin{equation} 
 \hat H_\mathrm{eff} = \hat H_0 +g \hat V_D+  
\frac{g^2}{2}\left[\hat S_0, \hat V_x\right].\label{effective-ham-proj-2}
\end{equation}

Substituting Eq.~\eqref{SO-operator} into Eq.~\eqref{effective-ham-proj-2}, we find
 \begin{multline} \label{LastTWOTERMs}
\left[\hat S_0, \hat V_X\right] = \left[\hat P \hat U_1 \hat Q + \hat Q\hat U_2\hat P, \hat P \hat V \hat Q + \hat Q \hat V \hat P\right]= \\ =
 \hat P\left[\hat U_1\hat Q\hat V-\hat V\hat Q\hat U_2\right]\hat P + \\+\hat Q\left[\hat U_2 \hat P \hat V - \hat V \hat P\hat U_1\right]\hat Q.
\end{multline}
Since the effective Hamiltonian does not mix the two subspaces $P$ and $Q$, one concludes that the second term on the {\it r.h.s.} of Eq.~\eqref{LastTWOTERMs} does not contribute to the projected effective Hamiltonian $\hat P\hat H_\mathrm{eff}\hat P$.
Taking that into account we finally obtain
the explicit matrix elements of the effective Hamiltonian of a well-defined group of quasi-degenerate states, $P$,
\begin{multline}
 \langle L_p|\hat H_\mathrm{eff}|R_{p^\prime}\rangle = E_p^{(0)}\delta_{p,p^\prime} + g\langle L_p|\hat V|R_{p^\prime}\rangle - \\ - \frac{g^2}{2}\sum_{q\in Q}\left(\frac{1}{E_q^{0}-E_p^{0}}+\frac{1}{E_q^{0}-E_{p^\prime}^{0}}\right)\times \\ \times\langle L_p|\hat V|R_q\rangle \langle L_q|\hat V|R_{p^\prime}\rangle.~~~~~~~~~~~~~~~~~\label{eff-ham-proj-final}
\end{multline}
Note, the outlined procedure is quite general and can be extended to obtain the higher-order corrections in $g$.

\section{Two $\BP\BT$-symmetric qubits embedded in a lossless resonator: Model and energy spectrum\label{section:model}}

To demonstrate the effectiveness of the non-hermitian version of the Schrieffer-Wolf transformation we consider a basic $\BP\BT$-symmetric circuit QED composed of two non-Hermitian qubits coupled to a lossless resonator. 
The Hamiltonian of such a system reads as
\begin{equation}
 \hat H  = \hat H_\mathrm{qb} + \hat H_\mathrm{res}+\hat H_\mathrm{int}, 
 \label{fullham}
 \end{equation}
where the Hamiltonian of two non-interacting biased qubits is
\begin{equation}
 \hat H_\mathrm{qb}  = \sum_{n=1}^2 \left [\frac{\Delta}{2}\hat\sigma_n^x+\frac{\epsilon}{2}\hat \sigma_n^z +(-1)^n i\gamma \hat \sigma_n^z \right ].\label{ham-qb}
 \end{equation}
Here, $\hat \sigma^{(x,z)}$ are the corresponding Pauli matrices, and 
$\Delta$, $\epsilon$ (a qubit's bias) are real off-diagonal and diagonal matrix elements, accordingly. The non-hermiticity is introduced as a \textit{staggered imaginary} longitudinal magnetic field or a gain/loss parameter of individual qubits, $\gamma$.
All parameters of individual qubits are assumed to be identical. 

The Hamiltonian $\hat H_\mathrm{res}$ describes a single bosonic mode of the resonator with the frequency $\omega_r$ 
 \begin{equation}
 \hat H_\mathrm{res}  = \omega_r\hat a^\dagger \hat a,\label{ham-res}
 \end{equation}
 where $\omega_r$ is the characteristic frequency of the resonator.
The interaction between the qubits and the resonator is determined by the Hamiltonian $\hat H_\mathrm{int}$ as
 \begin{equation}
\hat H_\mathrm{int}  = (\hat a^\dagger +\hat a)(g_1\hat\sigma_{1}^z+g_2\hat\sigma_2^z).\label{ham-int}
\end{equation}
Identical coupling strengths, $g=g_1=g_2$, are assumed for simplicity. 
%
%

 
The parity $\hat\BP$ and the time-reversal $\BT$ operators are defined for the system as the exchange of qubits and the complex conjugation operators, respectively~\cite{Song-14,tetling2022linear,starkov2023quantum}:
\begin{equation}
    \hat \BP \hat\sigma_j^r \hat \BP^{-1} = \hat \sigma_{3-j}^r, ~~~~
   \hat \BT i \hat \BT^{-1} = -i. \label{parity_timedef}
\end{equation}
Notice here that in the presence of the opposite
signs of the gain/loss parameter for different qubits, {\it i.e.}, a
staggered gain/loss, the $\BP\BT$ symmetry of the Hamiltonian
\eqref{fullham} is preserved even in the biased regime, $\epsilon \neq 0$. In addition to the $\BP\BT$ symmetry, the Hamiltonian~\eqref{fullham} is also $\BP$-pseudo-Hermitian, i.e.,
\begin{equation}
    \hat \BP \hat H \hat \BP^{-1} = \hat H^\dagger,
\end{equation}
which will prove to be crucial for the qualitative understanding of the results.

Next, by making use of the direct numerical diagonalization we compute the low-lying eigenspectrum of the Hamiltonian \eqref{fullham}. The procedure described in detail in \cite{tetling2022linear,starkov2023quantum,PhysRevA.108.022206} is as follows: fixing the parameters $\Delta$ and $\epsilon$ we vary the gain (loss) $\gamma$ and the coupling strength, $g$. The maximum number of boson states used in the numerical analysis was $n=7$. Such numerically obtained eigenspectrum was then used to identify the $\BP\BT$--symmetry preserved and broken quantum phases, and the \textit{EP}s separating these phases.

Since the formation of \textit{EP}s in the model is based on  
the parity-indices of the states due to its $\BP$-pseudo-Hermitian symmetry~\cite{starkov2023quantum,PhysRevA.108.022206}, we 
assign to each level a topological index equal to the parity of the level at $\gamma=0$~\footnote{Note that $\BP$ pesudo-Hermiticity implies that $\BP$ commutes with the Hamiltonian at $\gamma=0$.}. This index is conserved in the region of parameters where an eigenvalue of the state stays real, and the second-order \textit{EP}s are provided only by pairs of states with opposite parity-indices \cite{PhysRevA.108.022206}.

\subsection{Hermitian circuit QED, $\gamma=0$: longitudinal and transverse couplings}

\begin{figure*}[t]
\begin{center}
\includegraphics[width=450pt]{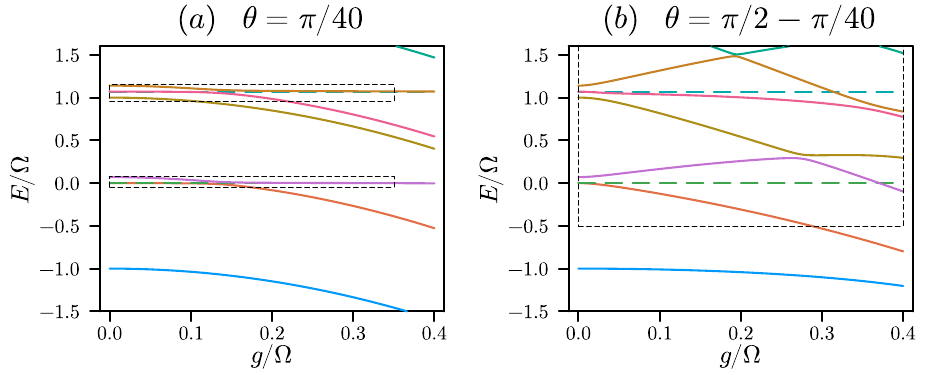}
\caption{Calculated lowest-lying energy levels of the  Hermitian circuit QED ($\gamma=0$) biased in the resonant regime. The mixing angle was chosen as: $a$) $\theta= \pi/40 = 4.5^\circ$ providing an almost longitudinal coupling; $b$)  $\theta= \pi/2-\pi/40 = 85.5^\circ$ providing an almost transverse coupling.
Solid (dashed) curve denotes the even (odd) parity of the levels, respectively.
The dashed rectangles show the parts of the spectrum that are zoomed in Figs.~\ref{fig:group1} (middle one of panel $(a)$), ~\ref{fig:group2} (upper one of panel $(a)$), and \ref{fig:transversal} (the single one of panel $(b)$).}
\label{fig:overview-1}
\end{center}
\end{figure*}

First, we recall the quantum-mechanical behavior of a basic circuit QED system in the absence of non-Hermitian terms, i.e., as $\gamma=0$ \cite{blais2004cavity,krantz2019quantum}. In this case one can define
the frequency $\Omega$ and the mixing angle $\theta$ of a single qubit as 
\begin{equation}
\Omega = \sqrt{\Delta^2+\epsilon^2},\quad \theta = \arccos{\left(\frac{\epsilon}{\sqrt{\Delta^2+\epsilon^2}}\right)}.
\end{equation}
Transforming the single-qubit Hamiltonian $\hat{H}_{1qb} = (\Delta/2)\hat\sigma^x + (\epsilon/2)\hat\sigma^z$ to the diagonal form we obtain two eigenvectors $|\pm \rangle$ with corresponding eigenvalues $\pm \Omega/2$. In a new representation the interaction Hamiltonian  takes the form:
\begin{equation} \label{intHamiltonian-2}
\hat{\tilde H}_\mathrm{int} = g[\cos \theta (\hat\sigma_{1}^z+\hat\sigma_2^z)+\sin \theta (\hat\sigma_{1}^x+\hat\sigma_2^x)](\hat a^\dagger +\hat a).
\end{equation}
We choose the parameters to be close to the resonance condition, $|\omega_r-\Omega| \ll \Omega$,  precisely $\omega_r/\Omega = 1.07$. 
The numerically obtained nine low-lying normalized energy levels, $E/\Omega$, of Eq.~\eqref{fullham} (for $\gamma=0$) are presented in  Figs.~\ref{fig:overview-1}(a,b) as a function of normalized interaction strength~$g/\Omega$ for two extreme interacting regimes: almost \textit{longitudinal}
($\sin \theta \simeq 0$) and \textit{transverse} ($\cos \theta \simeq 0$) couplings, respectively.

The energy levels of a Hermitian circuit QED biased in the resonant regime demonstrate the groups of well-separated quasi-degenerate states for $g \ll 1$, and therefore, the Schrieffer-Wolf transformation are suitable for the analysis of such a system. We also relate the energy levels plotted in Fig.~\ref{fig:overview-1} to the eigenvectors at $g=0$. In the order of the increasing energy the ground state corresponds to
\begin{equation}
|--0\rangle,
\end{equation}
the middle group of three states corresponds to
\begin{gather}
\frac{|+-\rangle+|-+\rangle}{\sqrt{2}}\otimes |0\rangle,\quad \frac{|+-\rangle-|-+\rangle}{\sqrt{2}}\otimes |0\rangle,\nonumber \\ |--1\rangle,\label{group1-real}
\end{gather}
and the upper group of four states is
\begin{gather}
    |++0\rangle,\quad \frac{|+-\rangle+|-+\rangle}{\sqrt{2}}\otimes |1\rangle,\nonumber\\
    \frac{|+-\rangle-|-+\rangle}{\sqrt{2}}\otimes |1\rangle,\quad |--2\rangle.\label{group2-real}
\end{gather}
Here, the notation implies $|\pm\pm n\rangle = |\pm_1\rangle\otimes|\pm_2\rangle\otimes |n\rangle$, where $n$ is the number of bosons in the resonator.
The parities of all these states are positive ($+1$)  except of the states
\begin{equation}
\frac{|+-\rangle-|-+\rangle}{\sqrt{2}}\otimes |0\rangle,\quad \frac{|+-\rangle-|-+\rangle}{\sqrt{2}}\otimes |1\rangle 
\end{equation}
which have negative ($-1$) parity. 
According to the Wigner-von Neumann selection rule~\cite{NeumannWigner-29}, the latter states are decoupled from the rest of the states. For this reason, these negative parity states are called "dark" while the rest of the states are referred to as "bright". However, at non-zero $\gamma$ the non-Hermitian terms in the 
$\BP$-pseudo-Hermitian Hamiltonian mix the "dark" and the "bright" states resulting in the second-order EPs \cite{PhysRevA.108.022206}.

\subsection{Eigenspectrum of $\BP\BT$-symmetric circuit QED, $\gamma \neq 0$: longitudinal coupling}

\begin{figure*}[t]
\begin{center}
\includegraphics[width=450pt]{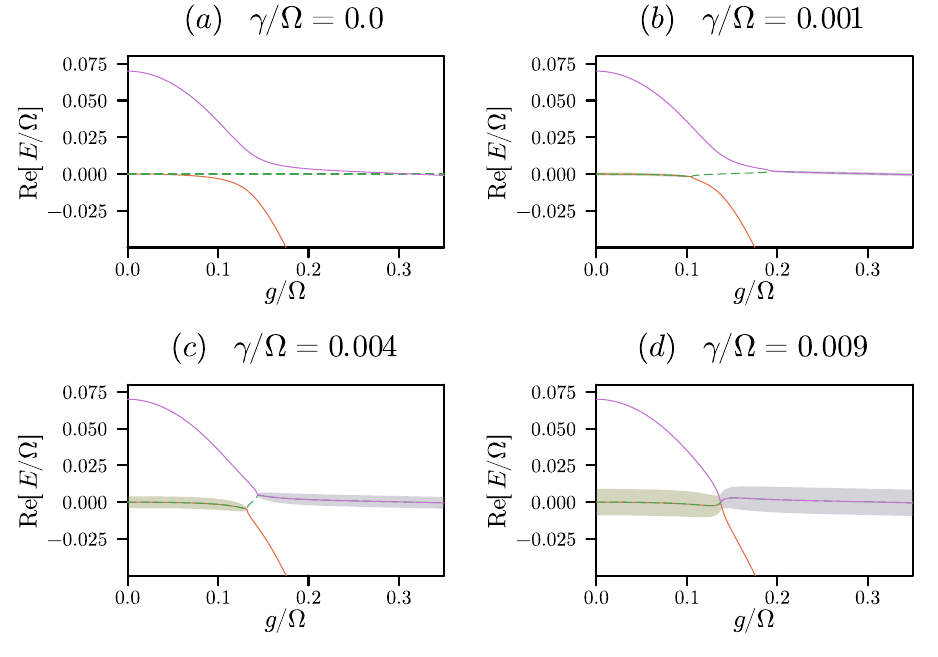}
\caption{Calculated real parts $\re{\left[E/\Omega\right]}$ of the eigenenergies of the quasi-degenerate states as functions of the normalized interaction strength $g/\Omega$ for different values of $\gamma/\Omega$. The panels zoom on the three middle levels of Fig.~\ref{fig:overview-1}(a).
The widths of the shaded ribbons are proportional to the imaginary parts $\im{\left[E/\Omega\right]}$ of the eigenenergies. Solid (dashed) lines correspond to positive (negative) parity of the states at $\gamma=0$.
The qubit mixing angle $\theta$ was taken the same as for Fig.~\ref{fig:overview-1}(a).}
\label{fig:group1}
\end{center}
\end{figure*}

\begin{figure*}[t]
\begin{center}
\includegraphics[width=450pt]{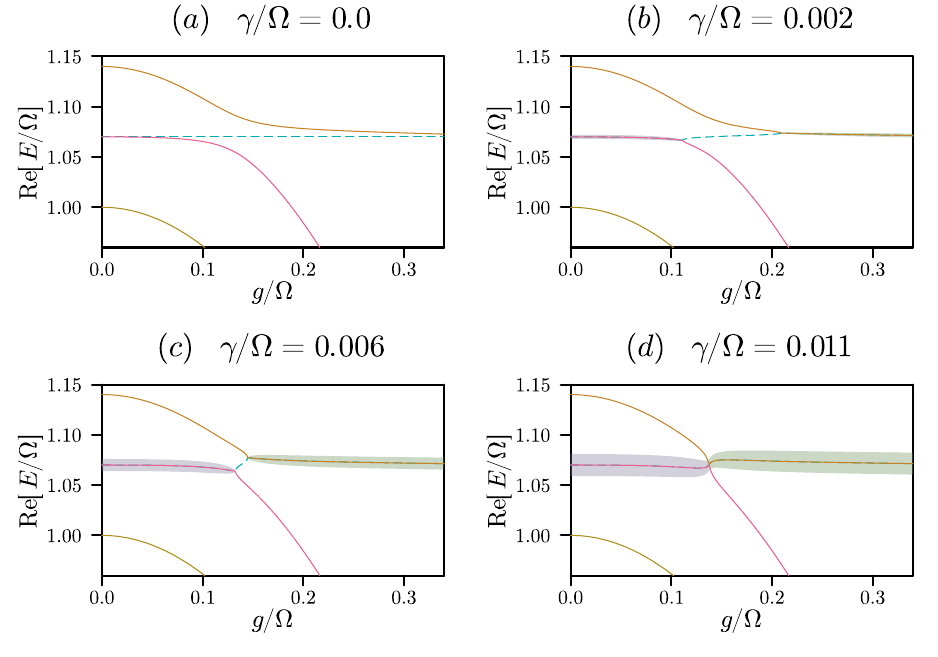}
\caption{Calculated real parts $\re{\left[E/\Omega\right]}$ of the eigenenergies  of the quasi-degenerate states as functions of the normalized interaction strength $g/\Omega$ for various values of $\gamma/\Omega$. The panels zoom on the four upper levels of Fig.~\ref{fig:overview-1}a. The widths of the shaded ribbons are proportional to the imaginary parts $\im{\left[E/\Omega\right]}$ of the eigenenergies. Solid (dashed) lines correspond to positive (negative) parity of the states at $\gamma=0$. The 
qubit mixing angle is the same as for Fig.~\ref{fig:overview-1}(a).}
\label{fig:group2}
\end{center}
\end{figure*}

To study the eigenspectrum of $\BP\BT$-symmetric circuit QED with an almost longitudinal coupling ($\theta=\pi/40 \ll 1$) between qubits and the resonator it is convenient to focus separately on the two  groups of quasi-degenerate states that we identified in Fig.~\ref{fig:overview-1}(a). In Figs.~\ref{fig:group1} and~\ref{fig:group2}, we display the real parts of the normalized eigenenergies, $\re\left[E/\Omega\right]$, as a function of the normalized interaction strength, $g/\Omega$, for different fixed values of $\gamma/\Omega$. The positive (negative) parities of the states are indicated by solid (dashed) lines.
In Fig.~\ref{fig:group1}, we focus on the group of three middle levels of Fig.~\ref{fig:overview-1}(a), while in Fig.~\ref{fig:group2} we focus on the group of four upper levels.
In the parameter regions where we have pairs of complex conjugated eigenvalues, we display the imaginary parts $\im\left[E/\Omega\right]$ of the eigenenergies as the shaded ribbons with the widths proportional to $\im\left[E/\Omega\right]$. A pair of complex conjugated eigenvalues can appear, when two real-valued levels get together and pass through a second-order \textit{EP}. Therefore, we identify the second-order \textit{EP}s in Figs.~\ref{fig:group1} and~\ref{fig:group2} as the points where the ribbon width shrinks to zero. As expected~\cite{PhysRevA.108.022206}, the second-order \textit{EP}s are formed between the states with opposite parity indices.

The panels $(b,c)$ of Fig.~\ref{fig:group1} demonstrate the precursor of a third-order~\textit{EP}.  One finds two second-order \textit{EP}s separated by the region of parameters where the three levels have real eigenenergies. Moreover, these two second-order \textit{EP}s share only one level with opposite parity index. As $\gamma/\Omega$ increases, the intermediate region with three real eigenvalues shrinks until two \textit{EP}s coalesce into one point, which is the third-order \textit{EP}. For a resonant $\BP\BT$-symmetric circuit QED with an almost longitudinal coupling a third-order \textit{EP} occurs for an extremely small value of 
$\gamma/\Omega \simeq 0.008$.

The three upper levels of Fig.~\ref{fig:group2} also form a third-order \textit{EP}. We show its precursor in panel $(c)$ of Fig.~\ref{fig:group2}. To conclude this subsection we notice that the higher energy levels (not shown in Fig.~\ref{fig:overview-1})  also form well-separated groups of four quasi-degenerate levels that behave similarly to the one considered in Fig.~\ref{fig:group2}.

\subsection{Eigenspectrum of $\BP\BT$-symmetric circuit QED, $\gamma \neq 0$: transverse coupling}

\begin{figure*}[t]
\begin{center}
\includegraphics[width = 450pt]{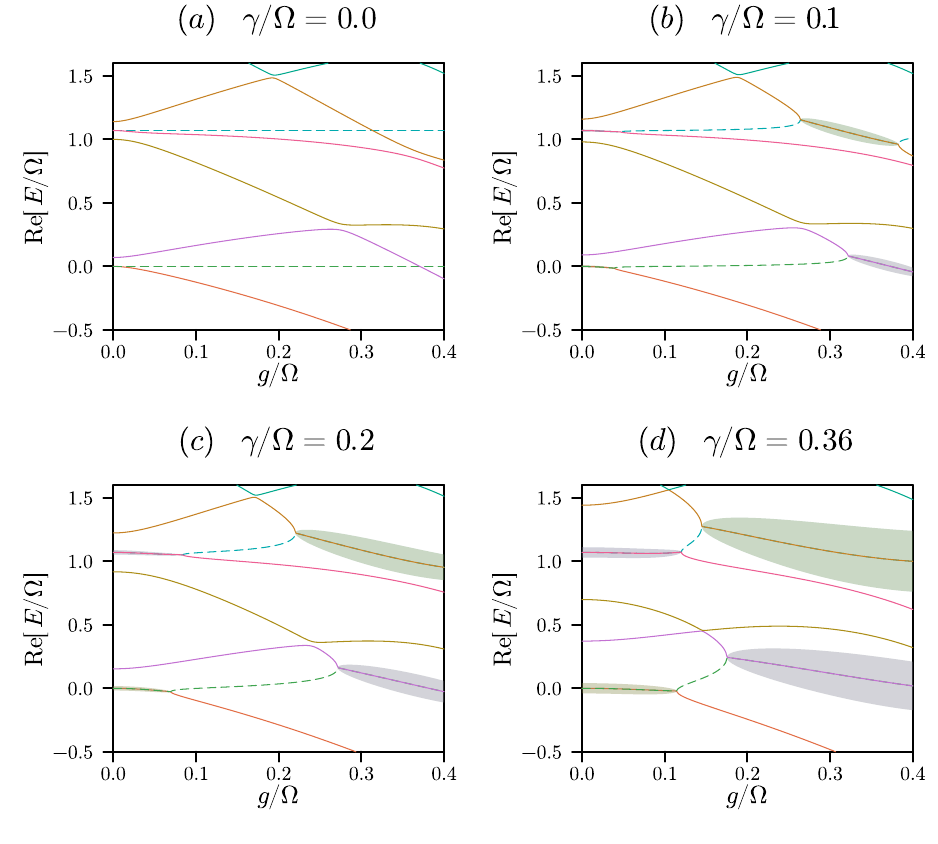}
\caption{Calculated real parts $\re{\left[E/\Omega\right]}$ of the eigenenergies  of the lowest-lying levels as functions of the normalized interaction strength $g/\Omega$ for various values of $\gamma/\Omega$. The widths of the shaded ribbons are proportional to the imaginary parts $\im{\left[E/\Omega\right]}$ of the eigenenergies. The resonator frequency is chosen $\omega_r/\Omega=1.07$. The qubit mixing angle $\theta=\pi/2-\pi/40$. Solid (dashed) lines correspond to positive (negative) parity of the states at $\gamma=0$.}
\label{fig:transversal}
\end{center}
\end{figure*}

In Fig.~\ref{fig:transversal}, we present the eigenspectrum of $\BP\BT$-symmetric circuit QED biased in the resonant regime 
in the presence of an almost transverse coupling for the qubit mixing angle, $\theta=\pi/2-\pi/40$, and different values of $\gamma/\Omega$.
Similarly to the previous subsection we focus on the groups of excited levels as in Figs.~\ref{fig:group1} and~\ref{fig:group2}, but display them together in this case.

Qualitative picture roughly does not change: the non-hermiticity leads to the formation of the second-order \textit{EP}s between the levels with opposite parity indices. We also observe pairs of second-order \textit{EP}s with shared level of opposite parity index that approach each other as $\gamma$ increases. These pairs eventually coalesce into third-order \textit{EP}s.

Nevertheless, there is also a significant difference with the case of almost longitudinal coupling: as the qubit-resonator interaction $g$ increases, the level repulsion between the states of the same parity index inside a resonant group is so strong, that it pushes the states from different groups towards each other and they go through avoided crossings. Therefore, the applicability of the Schrieffer-Wolf transformation for the $\BP\BT$--symmetric circuit QED with a transverse coupling is limited to the regime $g/\Omega \ll 1$.

\section{Two $\BP\BT$--symmetric circuit QED: Effective Hamiltonian of  a quasi-degenerate group of states.\label{section:effham}}

Here, we obtain analytically the eigenspectrum of $\BP\BT$-symmetric circuit QED by making use of the generic Schrieffer-Wolf transformation elaborated in Sec. II. We identify 
the main part $\hat H_0$ of the Hamiltonian \eqref{ham-perturb} as 
$\hat H_0= \hat H_\mathrm{qb} + \hat H_\mathrm{res}$, and the perturbation $g\hat V=\hat H_\mathrm{int}$.
We focus on a $\BP\BT$-symmetric circuit QED with a longitudinal coupling and on the quasi-degenerate group of states presented in Fig.~\ref{fig:group1}.

\subsection{Biorthonormal basis of the right and the left eigenvectors at $g=0$.}

Let us first discuss the eigenvalues and eigenvectors of unperturbed $\BP\BT$-symmetric circuit QED, i.e., as $g=0$. The eigenvalues of the single-qubit Hamiltonian $\hat H_{1\mathrm{qb}}=\left(\Delta/2\right)\hat\sigma_1^x+\left(\epsilon/2+i\gamma\right)\hat\sigma_1^z$ are readily obtained as
\begin{equation}
 \varepsilon_\pm = \pm \lambda = \pm\sqrt{\left(\frac{\Delta}{2}\right)^2+\left(\frac{\epsilon}{2}+i\gamma\right)^2}.\label{lambda-def}
\end{equation}
The corresponding  right eigenvectors are chosen as
\begin{align}
 |+_r\rangle & = \begin{pmatrix}
                \frac{\epsilon}{2}+i\gamma+\lambda\\ \frac{\Delta}{2}
               \end{pmatrix},\\
 |-_r\rangle & = \begin{pmatrix}
                -\frac{\Delta}{2}\\ \frac{\epsilon}{2}+i\gamma+\lambda
               \end{pmatrix}.
\end{align}
As $\hat H_{1\mathrm{qb}}$ is described by a symmetric matrix, the corresponding left eigenvectors are obtained by the transposition:
\begin{align}
 \langle +_l| & = \frac{1}{\left(\frac{\Delta}{2}\right)^2 + \left(\frac{\epsilon}{2}+i\gamma+\lambda\right)^2} \begin{pmatrix}
              \frac{\epsilon}{2}+i\gamma+\lambda, & \frac{\Delta}{2}
             \end{pmatrix},\label{lplus}\\
 \langle -_l| & = \frac{1}{\left(\frac{\Delta}{2}\right)^2 + \left(\frac{\epsilon}{2}+i\gamma+\lambda\right)^2} \begin{pmatrix}
             -\frac{\Delta}{2}, & \frac{\epsilon}{2}+i\gamma+\lambda
             \end{pmatrix}.\label{lminus}
\end{align}
The chosen right and left eigenvectors produce the biorthonormal basis states of a single qubit. Indeed, one can check that 
$\langle +_l|+_r\rangle=\langle -_l|-_r\rangle=1$, and  
$\langle +_l|-_r\rangle=\langle -_l|+_r\rangle=0$. 

The Hamiltonian of the second qubit $\hat H_{2\mathrm{qb}}$ is obtained from $\hat H_{1\mathrm{qb}}$ by complex conjugation. As the consequence, the eigenvalues and both right and left eigenvectors can be obtained by the complex conjugation of the expressions (\ref{lambda-def})-(\ref{lminus}). 
%
At $g=0$ the vectors
\begin{equation}
|R_{\pm\tilde\pm n}\rangle =|\pm_r\rangle\otimes|\tilde\pm_r\rangle\otimes|n\rangle\label{rightbasis}
\end{equation}
constitute the basis of the right eigenvectors of the Hamiltonian $\hat H_0$ with eigenenergies
\begin{equation}
E_{\pm\pm n} = (\pm\lambda\pm\lambda^*)+n\omega_0.
\end{equation}
Here, $\sim$ in $|\tilde \pm_r\rangle$ serves as a reminder that the eigenenergies and the eigenvectors of the second qubit are obtained by complex conjugation of (\ref{lambda-def})-(\ref{lminus}). 
The corresponding basis of the left eigenvectors is
\begin{equation}
\langle L_{\pm\tilde\pm n}|=\langle\pm_l|\otimes\langle\tilde\pm_l|\otimes\langle n|.\label{leftbasis}
\end{equation}
Notice that the normalized boson states $|n\rangle$ are simultaneously the right and the left eigenvectors, so they do not have any $r/l$ subscripts. Thus defined eigenvectors~\eqref{rightbasis} and~\eqref{leftbasis} form the complete biorthonormal basis of the Hamiltonian $H_0$.

For $\gamma\ll\Omega$ one finds $\lambda\approx \Omega$. For resonant $\BP\BT$-symmetric circuit QED, i.e., as $\omega_r\approx \Omega$, the following states form the group of quasi-degenerate levels (the last state is absent for $n=0$):
\begin{equation*}
 |+_r\tilde-_r;~ n\rangle,\quad |-_r\tilde+_r;~ n\rangle,
\end{equation*}
\begin{equation}
 |-_r\tilde-_r; ~(n+1)\rangle,\quad |+_r\tilde+_r; ~(n-1)\rangle.
\end{equation}
For $\gamma=0$ these states transform in Eqs.~\eqref{group1-real} and~\eqref{group2-real}. 


\subsection{Effective Hamiltonian for a quasi-degenerate group of eigenstates}

Here, using the main result (\ref{eff-ham-proj-final}) we obtain the effective Hamiltonian for the group of quasi-degenerate states with $n=0$. Then the set $P$ of states that we want to project onto contains only three states: the right eigenvectors are
\begin{align}
 |R_1\rangle & = |+_r \tilde-_r; ~0\rangle,\nonumber\\
 |R_2\rangle & =|-_r \tilde+_r ;~0\rangle,\label{prbasis}\\
 |R_3\rangle & =|-_r \tilde-_r; ~1\rangle,\nonumber
\end{align}
and the left eigenvectors are
\begin{align}
 \langle L_1| & = \langle+_l \tilde-_l;~0|,\nonumber\\
 \langle L_2| & =\langle-_l \tilde+_l;~ 0|,\label{plbasis}\\
 \langle L_3| & =\langle-_l \tilde-_l;~ 1|\nonumber
\end{align}
The corresponding energies at $g=0$ are
\begin{gather}
 E_1^{(0)} = \lambda-\lambda^*,\quad E_2^{(0)} = \lambda^*-\lambda,\nonumber\\
 E_3^{(0)} = \omega_0 - (\lambda+\lambda^*).\label{g0-energies}
\end{gather}

The evaluation of the direct matrix elements of the pertubration $\langle L_p|\hat V|R_{p^\prime}\rangle$ and of the second-order contribution due to indirect transitions in Eq.~\eqref{eff-ham-proj-final} is straightforward but lengthy, so we provide it in Appendix~\ref{section:matelems}.

The explicit expression for the effective Hamiltonian $\hat H_\mathrm{eff}$ is given by
\begin{widetext}
\begin{multline}
\langle L_p|\hat H_\mathrm{eff}| R_{p^\prime}\rangle=\\
 \begin{pmatrix}
  2i\im\lambda - g^2\left[\frac{(t^*)^2}{\omega_0+2\lambda^*} - \frac{4(\im s)^2}{\omega_0}\right] & -g^2|t|^2\re\left(\frac{1}{\omega_0+2\lambda}\right) & -gt \\
  -g^2|t|^2 \re\left(\frac{1}{\omega_0+2\lambda}\right) & -2i\im\lambda - g^2\left[\frac{t^2}{\omega_0+2\lambda} - \frac{4(\im s)^2}{\omega_0}\right] & -gt^*\\
-gt & -gt^* & \omega_0 - 2\re \lambda - 4g^2\left[\frac{(\re s)^2}{\omega_0} + \re\left(\frac{t^2}{\omega_0+2\lambda}\right)\right]\label{eff-matrix-full}
 \end{pmatrix}
\end{multline}
\end{widetext}
Here, the parameters $s$ and $t$ are the single qubit matrix elements defined in Appendix~\ref{sspin}:
\begin{equation} \label{s-expression-1}
    s = \langle+_l|\hat\sigma_z|+_r\rangle = -\langle-_l|\hat\sigma_z|-_r\rangle,
\end{equation}
\begin{equation} \label{Texp1}
    t = -\langle+_l|\hat\sigma_z|-_r\rangle = -\langle-_l|\hat\sigma_z|+_r\rangle.
\end{equation}
The coefficient $s$ is the probability amplitude for the qubit to stay in the same state, while $t$ is the probability amplitude to flip the qubit.
Since $\gamma/\Omega \ll 1$, the expressions for $s$ and $t$ are simplified as 
\begin{equation} \label{s-expression}
 s = \frac{\epsilon/2+i\gamma}{\lambda} \approx \cos\theta +\frac{2i\gamma}{\Omega}\sin^2\theta,
\end{equation}
\begin{equation} \label{Texpression}
    t = \frac{\Delta}{2\lambda} \approx \sin{\theta} - \frac{i\gamma}{\Omega}\sin{2\theta},
\end{equation} 
and $\lambda \approx \Omega/2 + i\gamma\cos{\theta}$.

\begin{figure*}[t]
\begin{center}
\includegraphics[width=450pt]{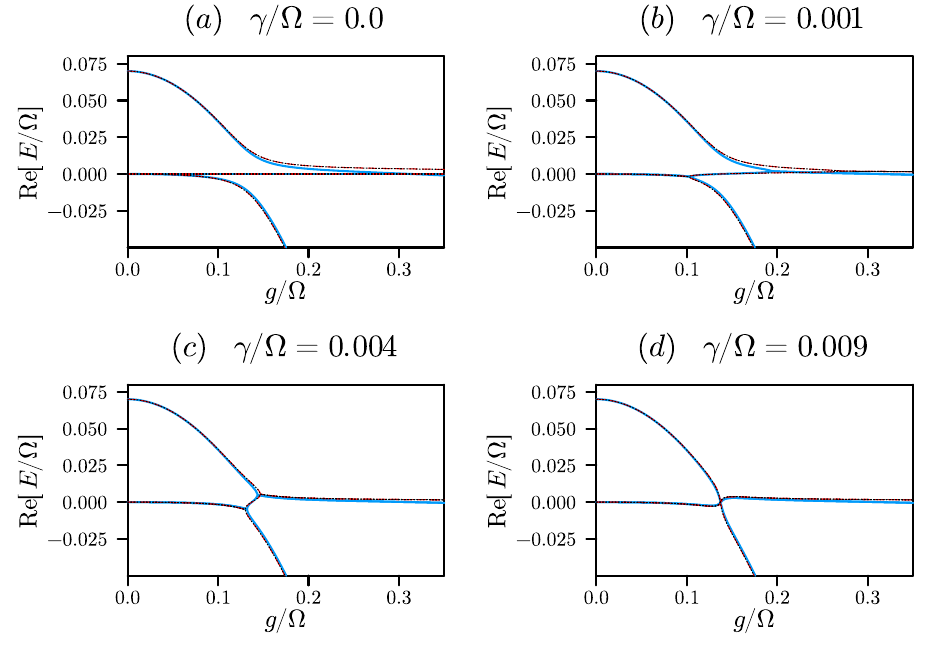}
\caption{Comparison of the calculated eigenenergies of the resonant states obtained by direct numerical diagonalization and from the effective Hamiltonian (\ref{eff-matrix-full}). Solid blue lines: real parts of the eigenenergies $\re\left[E/\Omega\right]$ obtained by direct numerical diagonalization of~\eqref{fullham}. Dashed red lines: the same, but obtained from the full effective Hamiltonian matrix~\eqref{eff-matrix-full}. Dotted black lines: the same, but obtained from the effective Hamiltonian matrix expanded to linear order in $\gamma/\Omega$. Resonator frequency ratio and the qubit mixing angle are the same as in Figs.~\ref{fig:overview-1}$(a)$, \ref{fig:group1} and \ref{fig:group2}: $\omega_r/\Omega=1.07$ and $\theta = \pi/40$.}
\label{fig:comparison}
\end{center}
\end{figure*}

To assess the quality of the generic Schrieffer-Wolf transformation and the effective Hamiltonian approximation, we compare in Fig.~\ref{fig:comparison} the real parts of the eigenenergies of the resonant states obtained in two different ways: by direct numerical diagonalization of~\eqref{fullham} and diagonalization of the effective Hamiltonian~\eqref{eff-matrix-full}. Observe that the energies obtained from the effective Hamiltonian start to slightly deviate around $g/\Omega \approx0.15$. However, the third-order \textit{EP} occurs at small enough $g/\Omega$ that is within the region of the validity of the approximation. We shall also point out, that tuning the resonator frequency closer to the qubit's frequency moves the position of the third-order \textit{EP} to smaller values of $g/\Omega$, and, therefore, makes the approximation to work better (see the Sec. \ref{section:effham-analysis}A for details).

The reasons of such a slight disagreement  
can be understood in the following way. The denominators in Eq.~\eqref{eff-ham-proj-final} contain the energy differences at $g=0$. When $g$ increases the energies deviate from their values at $g=0$, but as long as these deviations are small in comparison with energy differences $E_p^{(0)}-E_q^{(0)}$, we can neglect the renormalization of the energies in Eq.~\eqref{eff-ham-proj-final}. As one can see in Fig. \ref{fig:overview-1}, at relatively small values of $g$, the level repulsion brings the levels from different resonant groups sufficiently close in energy, so that a main assumption of a well-defined groups of quasi-degenerate states becomes invalid. Correspondingly, the effective Hamiltonian approximation can be improved if the energies entering Eq.~\eqref{eff-ham-proj-final} were to be determined in a self-consistent manner.

\subsection{Parity analysis of the effective Hamiltonian\label{section:parity}}

In order to provide the parity analysis of $\BP\BT$-symmetric QED, we notice that an exchanging of two qubits corresponds to the parity operator $\hat \BP$ in the subspace $P$ written as
\begin{equation}
 \hat\BP = \begin{pmatrix}
            0& 1&0\\
            1&0&0\\
            0&0&1
           \end{pmatrix}\label{eff-parity}
\end{equation}
Using~\eqref{eff-matrix-full} and~\eqref{eff-parity} one can show that $\hat \BP\hat H_\mathrm{eff} \hat \BP^{-1}= \hat H^\dagger$
Therefore, the effective Hamiltonian is $\BP\BT$-symmetric and $\BP$-pseudo-Hermitian as well.

Using these properties of $\hat H_\mathrm{eff}$ 
we apply the parity analysis of Ref.~\cite{PhysRevA.108.022206} directly to the effective Hamiltonian  to qualitatively describe the structure of the emergent \textit{EP}s. 
We recall here that the levels of a $\BP$--pseudo-Hermitian system can be characterized by their parity indices in the regions of parameters where the corresponding energies stay real. These parity indices are simply the parities of the states in the absence of non-Hermiticity. The formation of second-order \textit{EP}s is possible then only between the levels with opposite parity indices, and the higher-order \textit{EP}s can be obtained by tuning several second-order \textit{EP}s to coalesce.

To facilitate such parity analysis, it is natural to switch to a new basis of right
\begin{align}
    |R^\prime_1\rangle & = \frac{|+_r\tilde-_r\rangle+|-_r\tilde+_r\rangle}{\sqrt{2}}\otimes |0\rangle,\nonumber\\
    |R^\prime_2\rangle & = \frac{|+_r\tilde-_r\rangle-|-_r\tilde+_r\rangle}{\sqrt{2}}\otimes |0\rangle,\label{prbasis-new}\\
    |R^\prime_3\rangle & = |-_r\tilde-_r 1\rangle\nonumber
\end{align}
and left
\begin{align}
    \langle L^\prime_1| & = \frac{\langle +_l\tilde-_l|+\langle -_l\tilde+_l|}{\sqrt{2}}\otimes \langle 0|,\nonumber\\
    \langle L^\prime_2| & = \frac{\langle +_l\tilde-_l|-\langle -_l\tilde+_l|}{\sqrt{2}}\otimes \langle 0|,\label{plbasis-new}\\
    \langle L^\prime_3| & = \langle -_l\tilde-_l1|\nonumber
\end{align}
 eigenvectors, in which the parity operator~\eqref{eff-parity} is diagonal:
 \begin{equation}
 \hat \BP^\prime = \begin{pmatrix}
                    1 & 0&0\\
                    0&-1&0\\
                    0&0&1
                   \end{pmatrix}.\label{eff-parity-new}
 \end{equation}

 The transformation from the basis~\eqref{prbasis} and~\eqref{plbasis} to the basis~\eqref{prbasis-new} and~\eqref{plbasis-new} is described by the matrix $\hat F$
 \begin{equation}
    \begin{pmatrix}
        |R_1^\prime\rangle & |R_2^\prime\rangle & |R_3^\prime\rangle
    \end{pmatrix} = 
    \begin{pmatrix}
        |R_1\rangle & |R_2\rangle & |R_3\rangle
    \end{pmatrix} \hat F,
 \end{equation}
\begin{equation}
    \begin{pmatrix}
        \langle L_1^\prime| \\ \langle L_2^\prime| \\ \langle L_3^\prime|
    \end{pmatrix} = \hat F^{-1}
    \begin{pmatrix}
        \langle L_1| \\ \langle L_2| \\ \langle L_3|
    \end{pmatrix},
 \end{equation}
where
\begin{equation} \label{s-matrix}
     \hat F = \begin{pmatrix}
           \frac1{\sqrt 2} & \frac1{\sqrt 2}& 0\\
           \frac1{\sqrt 2} & -\frac1{\sqrt 2} & 0\\
           0 & 0 & 1
          \end{pmatrix}.
\end{equation}

Using (\ref{s-matrix}) we obtain the effective Hamiltonian  in the new basis:

\begin{widetext}
\begin{multline}
\langle L^\prime_p |\hat H_\mathrm{eff}|R^\prime_{p^\prime}\rangle = F^{-1}_{p,p_1}\langle L_{p_1}|\hat H_\mathrm{eff}|R_{p_1^\prime}\rangle F_{p_1^\prime,p^\prime}= \\ 
=
\begin{pmatrix}
g^2\left[\frac{(\Im s)^2}{\omega_0} - \re{\left(\frac{t^2 + |t|^2}{\omega_0+2\lambda}\right)}\right] & 2i\im{\left(\lambda + \frac{g^2t^2}{2(\omega_0+2\lambda)}\right)} & -\sqrt{2}g\re{t}\\
2i\im{\left(\lambda + \frac{g^2t^2}{2(\omega_0+2\lambda)}\right)} &
g^2\left[\frac{4(\im{s})^2}{\omega_0} + \re{\left(\frac{|t|^2-t^2}{\omega_0+2\lambda}\right)}\right] & -\sqrt{2}ig\im{t}\\
-\sqrt{2} g\re{t} & -\sqrt{2}ig\im{t} & \omega_0 - 2\re{\lambda} - 4g^2\left[\frac{(\re{s})^2}{\omega_0}+\re{\left(\frac{t^2}{\omega_0+2\lambda}\right)}\right]
\end{pmatrix}\label{eff-matrix-transf-basis}
\end{multline}
\end{widetext}
Taking into account the approximate expressions for $s$, $t$ and $\lambda$, i.e., (\ref{s-expression}) and (\ref{Texpression}), the effective Hamiltonian can be greatly simplified to
\begin{multline}
\langle L^\prime_p |\hat H_\mathrm{eff}|R^\prime_{p^\prime}\rangle \approx \\
 \begin{pmatrix}
  -\frac{2g^2\sin^2\theta}{\omega_r+\Omega} & 2i\gamma\cos\theta & -\sqrt{2}g\sin\theta\\
  2i\gamma\cos\theta & 0 & 0\\
  -\sqrt{2}g\sin\theta & 0 & \Delta\omega - 4g^2\left[\frac{\cos^2\theta}{\omega_r} + \frac{\sin^2\theta}{\omega_r+\Omega}\right]
 \end{pmatrix},\label{eff-matrix-appr}
\end{multline}
where the detuning frequency $\Delta\omega = \omega_r-\Omega$ is introduced.

At $\gamma=0$ and $g=0$, the three states~\eqref{prbasis} have the parities $1$, $-1$ and $1$ respectively. We identify the first and the third states as the "bright" states, while the second one --- as the "dark" state.
At $\gamma=0$ ,
 the dark state is decoupled from the bright states and 
 the Hermitian off-diagonal matrix elements, $-\sqrt{2}g\sin{\theta}$, couple the two "bright" states driving the \textit{avoided crossing } between them.
 
 
As $\gamma$ and $g$ are turned on, the middle state (the dark state at $\gamma=0$) with negative parity can go through a second-order \textit{EP} with either of the other two states, but the first and the third states having the same (positive) parity, never form a second-order \textit{EP}.
This property is reflected in the structure of the matrix~\eqref{eff-matrix-appr}: the diagonal matrix elements are purely real, and the  non-Hermitian off-diagonal matrix elements $2i\gamma\cos{\theta}$ only couple the middle state with the other two.
 

The interplay of the avoided crossing between the "bright" states and of the mixing between the "dark" state and one of the "bright" states due to non-hermiticity
leads to the precursor behaviour that we observed in Fig.~\ref{fig:group1}$(b,c)$. At $g=0$, the  two lowest-lying states form a complex conjugated pair. However, when $g$ increases, the avoided crossing leads to the exchange of the eigenvectors of the two "bright" states (first and third)~\cite{NeumannWigner-29}. As a result, the second level forms a complex conjugated pair with the third level instead.





\section{Quasi-degenerate states of $\BP\BT$-symmetric circuit QED: $\BP\BT$--symmetry preserved and broken phases 
\label{section:effham-analysis}}
The eigenvalues spectrum of an arbitrary  $\BP\BT$-symmetric quantum system can be of two types, i.e., real or complex conjugate values  defining
$\BP\BT$-symmetry unbroken (preserved) and broken quantum phases, accordingly. The \textit{EP}s of different orders separate these phases. 
By making use of the effective Hamiltonian (\ref{eff-matrix-appr}) in the following we identify these phases and the lines of \textit{EP}s in the group of resonant levels of $\BP\BT$-symmetric circuit QED. 

\subsection{Phase diagram ($\gamma/\Omega-g/\Omega$) of $\BP\BT$-symmetric circuit QED}

The eigenvalues of the effective Hamiltonian~\eqref{eff-matrix-appr}
are obtained as the roots of its characteristic polynomial, $P(E)=det (E-\hat H)=0$ \cite{tetling2022linear,Bergholtz-21,Sayyad-22, Kunst-22}. Since $\hat H_\mathrm{eff}$ is a $3\times3$ matrix, the  characteristic polynomial, $P(E)$, is written as $P(E)=E^3 + bE^2 + cE + d$, where the coefficients $b,~ c,~ d$ are expressed through the matrix elements of~\eqref{eff-matrix-appr} denoted as $H_{p,p^\prime}$;
\begin{align}
    b & = -\tr H_{p,p^\prime},\label{bdef}\\
    c & = \frac{\left(\tr H_{p,p^\prime}\right)^2 - \tr H_{p,p^\prime}^2}{2},\label{cdef}\\
    d & = -\det H_{p,p^\prime}.\label{ddef}
\end{align}
Because of the $\BP\BT$-symmetry of $H_{p,p^\prime}$, the coefficients of the characteristic polynomial are all purely real~\cite{Hatsugai-21}.

Transforming the characteristic polynomial, $P(E)$, to the dimensionless depressed cubic form, we obtain
\begin{equation}
   P(\tilde E)=\tilde E^3 + 3p\tilde E + 2q,\label{ch-pol-depressed}
\end{equation}
where the dimensionless energy $\tilde E$ is
\begin{equation}
    \tilde E = \frac{1}{\Omega}\left(E-\frac{b}{3} \right)\label{shift}
\end{equation}
and the coefficients $p$ and $q$ are
\begin{equation}
p  = \frac{1}{3\Omega^2}\left(c - \frac{b^2}{3}\right) 
\end{equation}
\begin{equation}
    q = \frac{2b^3-9cb+27d}{54\Omega^3} 
\end{equation}


Using the standard analysis of a depressed cubic equation~\eqref{ch-pol-depressed} (see Appendix~\ref{section:ch-pol-sol} for more details), we find that the equality, $p^3+q^2=0$, determines the line of second-order \textit{EP}s, and a third-order \textit{EP} is determined by the condition $p=q=0$.

\begin{figure*}[t]
\begin{center}
\includegraphics[width=500pt]{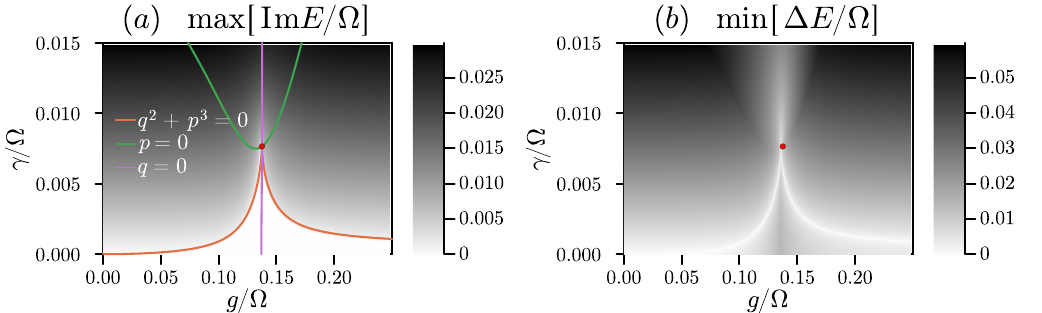}
\caption{Calculated maximal imaginary part of the eigenenergy $(a)$ and the minimal absolute value of the distance between two energies $(b)$ as the functions of the interaction strength $g/\Omega$ and gain (loss) parameter $\gamma/\Omega$. The line $p^3+q^2=0$ separates the regions of $\BP\BT$--symmetry unbroken (preserved) and broken phases (see red curve in the panel $(a)$). It also corresponds to a degeneracy in the eigenvalues (see panel $(b)$). The position of the third-order \textit{EP} is determined as the intersection of $p=0$ and $q=0$ lines (see panel $(a)$), and it also coincides with the fold in $p^3+q^2=0$ curve.}
\label{pq-lines}
\end{center}
\end{figure*}

The $\BP\BT$-symmetry preserved and broken phases can be quantitatively characterized by 
the maximal imaginary part of the eigenvalues of $\hat{H}_\mathrm{eff}$ 
and the minimal absolute value of the difference between two eigenvalues 
as the functions of parameters $g/\Omega$ and $\gamma/\Omega$. It is presented in Fig.~\ref{pq-lines}$(a,b)$.

The Fig.~\ref{pq-lines}$(a)$ allows us to distinguish the region of parameters, where $\BP\BT$--symmetry preserved phase occurs (white area),
from the region of parameters where $\BP\BT$--symmetry broken phase occurs (grayed area). The line $p^3+q^2=0$ (red curve) separates these two phases.
 The Figure~\ref{pq-lines}$(b)$ allows us to identify degeneracies of the eigenvalues. 
As we see, the line $p^3+q^2=0$ borders the region of parameters with purely real eigenvalues and corresponds to the degeneracy of the eigenvalues. As such, $p^3+q^2=0$ is simply the line of second-order \textit{EP}s.

The intersection of lines $p=0$ and $q=0$ (red point) corresponds to the triple-degenerate root of Eq.~\eqref{ch-pol-depressed} and indicates the third-order \textit{EP} at $g_\mathrm{cr}/\Omega \simeq 0.1375$  and $\gamma_{cr}/\Omega \simeq 7.65\times10^{-3}$. Notice here that the critical value of the interaction strength $g_\mathrm{cr}$ decreases with $\Delta \omega$ as $g_\mathrm{cr} \simeq \sqrt {(\Delta \omega)\Omega}/2$.
The critical value of the gain/loss parameter $\gamma_\mathrm{cr}$ satisfies the similar scaling, however, it acquires an additional degree of smallness due to almost longitudinal coupling $\theta\ll 1$: $\gamma_\mathrm{cr}\simeq g_\mathrm{cr}\times\theta/\sqrt{2}$ (see Appendix~\ref{section:critvals} for details).

In principle, triple-degeneracy of the characteristic polynomial root is not the sufficient condition to have a third-order \textit{EP}. However, we show explicitly in Appendix~\ref{section:ch-pol-sol}, that the triple-degeneracy point is indeed the third-order \textit{EP} in this case.

\subsection{The eigenvalues under perturbation of parameters $\gamma$ and $g$ near the third-order \textit{EP}.}

\begin{figure}[t]
\begin{center}
\includegraphics[width=225pt]{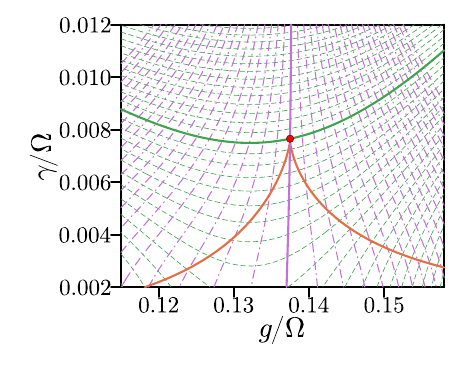}
\caption{The lines of constant $p$ and $q$ in the vicinity of the third-order \textit{EP}. Horizontally aligned green dashed lines: level curves $p=8\times10^{-6}\cdot m$ for integer $m$; solid line corresponds to the level $p=0$. Vertically aligned purple dash-dot lines: level curves $q = 2\times10^{-7}\cdot n$ for integer $n$; solid line corresponds to the level $q=0$. Orange solid line: level curve $p^3+q^2=0$.}
\label{pq-coords}
\end{center}
\end{figure}


To study the asymptotic behaviour of eigenvalues $E_{1-3}$ as the parameters $\gamma$ and $g$ slightly deviate from the values $\gamma_{cr}$ and $g_\mathrm{cr}$ corresponding to the third-order \textit{EP} ($EP_3$), it is convenient to work directly with the depressed cubic parameters $p$ and $q$ as a curvilinear system of coordinates in the vicinity of the third-order \textit{EP}. To corroborate this view, we display the lines of constant $p$ and $q$ in Fig.~\ref{pq-coords}.

Considering different directions of perturbation, we classify the behaviour of eigenvalues as follows (the details of the derivation can be found in Appendix~\ref{section:eigvals-perturb}).

Along the line of second-order \textit{EP}s (case 1), where the condition $p^3+q^2=0$ is valid, we obtain a single well-separated eigenvalue, $\tilde E_1 = -2\sqrt[3]{q}$ and doubly-degenerate eigenvalues, 
$\tilde E_{2,3} = -2\cos\frac{2\pi}{3}\times \sqrt[3]{q}$ (see Figs. \ref{fig:group1}c and \ref{fig:group2}c). The variation of parameters $\gamma$ and $g$ along the line $q=0, ~p<0$ (case 2) results in the \textit{real} spectrum with three distinct eigenvalues,  $\tilde E_k = 2\cos\left(\frac{\pi}{6}+\frac{2\pi}{3}(k-1)\right)\times\sqrt{|p|}$, where $k=1,2,3$. Such spectrum is also present in Figs. \ref{fig:group1}(c) and \ref{fig:group2}(c). The variation of parameters $\gamma$ and $g$ along the line $q=0, ~p>0$ (case 3) results in the  spectrum with a single $\tilde E_1=0$ and a pair of complex conjugated purely imaginary eigenvalues,  $\tilde E_{2-3} = \pm 2i\sin{\frac{2\pi}{3}}\times\sqrt{p}$ (see Fig. \ref{fig:group1}d and \ref{fig:group2}d). 

Finally, we notice that the variation of parameters $\gamma$ and $g$ along the line $q/p=const$ (case 4) leads into the $\BP\BT$--symmetry broken phase with the spectrum composed of a single real energy, $\tilde E_1 = -\sqrt[3]{2q}$ and two complex conjugated ones, $\tilde E_{2-3} = - e^{\pm 2i\pi/3}\sqrt[3]{2q}$. Such spectrum can be also seen in Figs. \ref{fig:group1}c,d and \ref{fig:group2}c,d.

Depending on the direction of the perturbation in the parameter space, we observe very different behaviour of the eigenvalues that an $EP_3$ splits into, i.e. anisotropic scaling of eigenvalues that was already reported for second-order \textit{EP}s~\cite{Ding-2018}.
It is worth noting, that in the previous works, the different behaviour of the eignevalues was attributed  to 
different types of $EP_3$. 
For example, the authors of Ref.~\cite{Sayyad-22} referred to the behaviour in cases $2-4$ as the $EP_3$ of types I, III and 0, respectively. In Ref.~\cite{Bergholtz-21}, the behaviour in the case $4$ was generically attributed to an $EP_3$ in a $\BP\BT$-symmetric system, while the behaviour in the case $2$ was generically attributed to a symmetric $EP_3$ in a $\BP$-symmetric system. 
Here we clearly demonstrate that the different behaviour of the eigenvalues under perturbation can be attributed to the very same $EP_3$.

\section{Conclusions 
\label{section:conclusions}}
To conclude, we extend the formalism of the generalized Schrieffer-Wolff transformation to arbitrary non-Hermitian interacting quantum systems, and derive the effective Hamiltonian taking into account the interaction terms as the perturbation up to the second order. To demonstrate the effectiveness of the method we apply this generic procedure to a basic $\BP\BT$--symmetric circuit QED composed of two non-Hermitian non-interacting superconducting qubits weakly coupled to a lossless resonator of the frequency $\omega_r$. In this model the non-Hermiticity is introduced as a staggered imaginary longitudinal magnetic field or gain/loss $\gamma$ of corresponding qubits states. We consider $\BP\BT$--symmetric circuit QED biased in the resonant regime as the frequency of a single qubit $\Omega$ is close to the frequency of the resonator, i.e., $\Omega  \simeq \omega_r$. The resonant regime of a circuit QED is a simplest one providing well-defined groups of quasi-degenerate eigenstates even for moderate values of the coupling  strength, $g/\Omega$.

Using the explicit expression for the effective Hamiltonian, we calculate analytically the eigenspectrum of a $\BP\BT$--symmetric circuit QED for different parameters $\gamma$ and $g$ and identify the $\BP\BT$--symmetry preserved and broken quantum phases, obtain the \textit{EP}s of different orders. In particular, we obtain the values of parameters $\gamma_{cr}$ and $g_\mathrm{cr}$ for the third-order \textit{EP}s. The values of $\gamma_{cr}$ and $g_\mathrm{cr}$ decrease with the detuning $\omega_r-\Omega$ and the mixing angle $\theta \ll 1$.

We compare our analytical results with the direct numerical diagonalization of the Hamiltonian (\ref{fullham}) and find a good agreement in the range of parameters $g, ~\gamma \ll 1$ for both a longitudinal and transverse coupling between qubits and the resonator. Moreover, for $\BP\BT$ quantum circuits with a \textit{longitudinal} coupling we find a good  agreement in a whole range of the interaction strength, $g/\Omega \leq 0.4$, and as a result one can obtain the position of the third-order \textit{EP} with a great accuracy.  The reason for that is the absence of a substantial mixing between eigenstates of different groups in $\BP\BT$--quantum circuit with a longitudinal coupling. 

The most important physical signature of the non-hermiticity is the mixture of "bright" and "dark" states. In particular, in the absence of non-Hermitian terms the energy spectrum of a circuit QED system splits into the subgroups of "bright" and "dark" states having different parity indices, and therefore, these subgroups do not mix with each other \cite{NeumannWigner-29}. As the non-Hermiticity is turned on, the second-order \textit{EP}s are formed between the pair of eigenstates with different parity (i.e. the "dark" and a one of "bright" states) due to $\BP$--pseudo-Hermitian symmetry of the Hamiltonian (\ref{fullham}). 
Moreover, the vicinity of the third-order \textit{EP} is characterized by the interplay of this non-Hermitian mixing and the standard avoided crossing between the "bright" states in the absence of dissipation.

Going beyond the analytical considerations of the present paper, it is important to emphasize that in the vicinity of an $n$-th order \textit{EP}, the states involved form an isolated quasi-degenerate group of states. As such, the Schrieffer-Wolff transformation outlined here can be used to numerically derive the \textit{local} $n\times n$ effective Hamiltonian describing the vicinity of the $n^{th}$-order \textit{EP}. Our formalism can also be readily applied to the thermodynamic models in condensed matter where Schrieffer-Wolf transformation is used to obtain the effective low-energy models by integrating out high-energy degrees of freedom.

\textbf{Acknowledgments }
We acknowledge the financial support of Deutsche Forschungsgemeinschaft (Projekt~EF~11/10-2) and the financial support through the European Union’s Horizon 2020 research and innovation program under grant agreement No 863313 'Supergalax'.

\appendix

\section{Derivation of the matrix elements of the effective Hamiltonian.\label{section:matelems}}

To derive the matrix elements of the effective Hamiltonian, we first need to compute the matrix elements of the perturbation:
\begin{multline}
    \langle L_{\pm\tilde\pm n}|\hat V |R_{\pm\tilde\pm m}\rangle = \\ =
    \langle \pm_l\tilde\pm_l;~ n|(\hat \sigma_1^z + \hat \sigma_2^z)(\hat a^\dagger+\hat a)|\pm_r\tilde\pm_r;~ m\rangle =\\ =
    \langle \pm_l\tilde\pm_l|(\hat \sigma_1^z + \hat \sigma_2^z)|\pm_r\tilde\pm_r\rangle\times\langle n|(\hat a^\dagger +\hat a)|m\rangle.
\end{multline}
The photon part of the matrix elements of $\hat V$ is obtained as:
\begin{equation}
\langle n|(\hat a^\dagger +\hat a)|m\rangle = \sqrt{n} \delta_{n,m+1} +\sqrt{n+1}\delta_{n+1,m}.
\end{equation}

The spin part of the matrix elements is computed in~\ref{section:spin-me}. After that in~\ref{section:eff-ham-comp}, we compute the direct and indirect contributions to the effective Hamiltonian obtaining Eq. (\ref{eff-matrix-full}).

\subsection{Spin part of the matrix elements\label{section:spin-me}}

Derivation of the spin part of the matrix elements proceeds in two steps.
In subsection~\ref{sspin}, we compute the single-spin matrix elements $\langle \pm_l |\hat \sigma^z|\pm_r\rangle$. In subsection~\ref{section:ts-me}, we use this single-spin matrix elements to compute the two-spin matrix elements $\langle \pm_l\tilde\pm_l|(\hat \sigma_1^z+\hat \sigma_2^z)|\pm_r\tilde\pm_r\rangle$.

\subsubsection{Single-spin matrix elements\label{sspin}}

Let us start with
\begin{multline}
 \langle+_l|\hat \sigma_z |+_r\rangle =\\
 \frac{1}{\left(\frac{\Delta}{2}\right)^2 + \left(\frac{\epsilon}{2}+i\gamma+\lambda\right)^2} \begin{pmatrix}
              \frac{\epsilon}{2}+i\gamma+\lambda, & \frac{\Delta}{2}
             \end{pmatrix}
             \begin{pmatrix}
  \frac{\epsilon}{2}+i\gamma+\lambda \\ -\frac{\Delta}{2}
 \end{pmatrix}
 =\\=
 \frac{1}{2\lambda\left(\frac{\epsilon}{2}+i\gamma+\lambda\right)}
 \begin{pmatrix}
  \frac{\epsilon}{2}+i\gamma+\lambda & \frac{\Delta}{2}
 \end{pmatrix}
 \begin{pmatrix}
  \frac{\epsilon}{2}+i\gamma+\lambda \\ -\frac{\Delta}{2}
 \end{pmatrix} = \\ =
 \frac{\left(\frac{\epsilon}{2}+i\gamma+\lambda\right)-\left(\frac{\Delta}{2}\right)^2}{2\lambda\left(\frac{\epsilon}{2}+i\gamma+\lambda\right)} = \\ = \frac{2\left(\frac{\epsilon}{2}+i\gamma\right)\left(\frac{\epsilon}{2}+i\gamma+\lambda\right)}{2\lambda\left(\frac{\epsilon}{2}+i\gamma+\lambda\right)} = \frac{\frac{\epsilon}{2}+i\gamma}{\lambda} = s,
\end{multline}
Here, we used the following two identities to factorize first the denominator and then the nominator:
\begin{multline}
 \left(\frac{\Delta}{2}\right)^2 + \left(\frac{\epsilon}{2}+i\gamma +\lambda\right)^2 =  \left(\frac{\Delta}{2}\right)^2 + \left(\frac{\epsilon}{2}+i\gamma\right)^2 + \\ +\left[\left(\frac{\Delta}{2}\right)^2 + \left(\frac{\epsilon}{2}+i\gamma\right)^2\right] + 2\lambda\left(\frac{\epsilon}{2}+i\gamma\right) = \\ =2\lambda\left(\frac{\epsilon}{2}+i\gamma+\lambda\right).
\end{multline}
\begin{multline}
\left(\frac{\epsilon}{2}+i\gamma +\lambda\right)^2 - \left(\frac{\Delta}{2}\right)^2 = \left(\frac{\epsilon}{2}+i\gamma\right)^2 +\\ + \left[\left(\frac{\Delta}{2}\right)^2 +\left(\frac{\epsilon}{2}+i\gamma\right)^2\right] + 2\left(\frac{\epsilon}{2}+i\gamma\right)\lambda-\left(\frac{\Delta}{2}\right)^2 = \\ =2\left(\frac{\epsilon}{2}+i\gamma\right)\left(\frac{\epsilon}{2}+i\gamma+\lambda\right)
\end{multline}

After that, we proceed with the other matrix elements in a straightforward manner.
\begin{multline}
 \langle-_l|\hat \sigma_z |-_r\rangle =\\ =
 \frac{1}{2\lambda\left(\frac{\epsilon}{2}+i\gamma+\lambda\right)}
 \begin{pmatrix}
  -\frac{\Delta}{2} & \frac{\epsilon}{2}+i\gamma+\lambda
 \end{pmatrix}
 \begin{pmatrix}
  -\frac{\Delta}{2} \\ -\left(\frac{\epsilon}{2}+i\gamma+\lambda\right)
 \end{pmatrix} = \\ =
 -\frac{\left(\frac{\epsilon}{2}+i\gamma+\lambda\right)-\left(\frac{\Delta}{2}\right)^2}{2\lambda\left(\frac{\epsilon}{2}+i\gamma+\lambda\right)} =  -s,
\end{multline}
\begin{multline}
 \langle+_l|\hat \sigma_z |-_r\rangle = \\= \frac{1}{2\lambda\left(\frac{\epsilon}{2}+i\gamma+\lambda\right)}
 \begin{pmatrix}
  \frac{\epsilon}{2}+i\gamma+\lambda & \frac{\Delta}{2}
 \end{pmatrix}
 \begin{pmatrix}
  -\frac{\Delta}{2} \\ -\left(\frac{\epsilon}{2}+i\gamma+\lambda\right)
 \end{pmatrix}= \\ =
 -\frac{\Delta\left(\frac{\epsilon}{2}+i\gamma+\lambda\right)}{2\lambda\left(\frac{\epsilon}{2}+i\gamma+\lambda\right)} = -\frac{\Delta}{2\lambda} = -t,
\end{multline}
\begin{multline}
 \langle-_l|\hat \sigma_z |+_r\rangle = \\ =\frac{1}{2\lambda\left(\frac{\epsilon}{2}+i\gamma+\lambda\right)}
 \begin{pmatrix}
  -\frac{\Delta}{2} & \frac{\epsilon}{2}+i\gamma+\lambda
 \end{pmatrix}
 \begin{pmatrix}
  \frac{\epsilon}{2}+i\gamma+\lambda \\ -\frac{\Delta}{2}
 \end{pmatrix} = \\ = -\frac{\Delta}{2\lambda} = -t.
\end{multline}

\subsubsection{Action of $\hat \sigma_1^z+\hat\sigma_2^z$ on two-spin states and the two-spin matrix elements.\label{section:ts-me}}

Here, we use the single-spin matrix elements obtained in the previous section. Instead of directly computing the two-spin matrix elements, it is more convenient to compute the action of $(\hat \sigma_1^z+\hat\sigma_2^z)$ on the two-spin states. The coefficients of the expansion then give the desired matrix elements.

\begin{multline}
 (\hat \sigma_1^z+\hat\sigma_2^z)|+_r\tilde-_r\rangle = (s|+_r\rangle - t|-_r\rangle)\otimes|\tilde -_r\rangle +\\+ |+_r\rangle\otimes(-s^*|\tilde-_r\rangle-t^*|\tilde+_r\rangle)
 =\\=
 (s-s^*)|+_r\tilde-_r\rangle - t|-_r\tilde-_r\rangle-t^*|+_r\tilde+_r\rangle,
\end{multline}
\begin{multline}
 (\hat \sigma_1^z+\hat\sigma_2^z)|-_r\tilde+_r\rangle = -(s|-_r\rangle + t|+_r\rangle)\otimes |\tilde+_r\rangle +\\+ |-_r\rangle\otimes(s^*|\tilde+_r\rangle-t^*|\tilde-_r\rangle)
 =\\=
 -(s-s^*)|-_r\tilde+_r\rangle - t|+_r\tilde+_r\rangle - t^*|-_r\tilde-_r\rangle,
\end{multline}
\begin{multline}
 (\hat \sigma_1^z+\hat\sigma_2^z)|-_r\tilde-_r\rangle = -(s|-_r\rangle - t|+_r\rangle)\otimes|\tilde-_r\rangle -\\- |-_r\rangle\otimes(s^*|\tilde-_r\rangle-t^*|\tilde+_r\rangle)
 =\\=
 -(s+s^*)|-_r\tilde-_r\rangle - t|+_r\tilde-_r\rangle - t^*|-_r\tilde+_r\rangle,
\end{multline}
\begin{multline}
 (\hat \sigma_1^z+\hat\sigma_2^z)|+_r\tilde+_r\rangle = (s|+_r\rangle - t|-_r\rangle)\otimes |\tilde+_r\rangle +\\+ |+_r\rangle\otimes(s^*|\tilde+_r\rangle-t^*|\tilde-_r\rangle)
 =\\=
 (s+s^*)|+_r\tilde+_r\rangle - t|-_r\tilde+_r\rangle - t^*|+_r\tilde-_r\rangle.
\end{multline}

As we have mentioned, the coefficients of the expansions in the right-hand sides gives us the matrix elements that we want.
For example,
\begin{equation}
 \langle +_l\tilde-_l|(\hat \sigma_1^z+\hat\sigma_2^z)|+_r\tilde-_r\rangle = (s-s^*).
\end{equation}

Notice that the matrix elements are symmetric with respect to the exchange of the initial and the final states.

\subsection{Matrix elements of the effective Hamiltonian for $n=0$\label{section:eff-ham-comp}}.

\subsubsection{The direct matrix elements of the perturbation.}

We first start we the second term in Eq.~\eqref{eff-ham-proj-final} for the effective Hamiltonian.
\begin{multline}
 \langle+_l \tilde-_l 0|\hat V |-_r \tilde-_r 1\rangle = \langle-_l \tilde-_l 1|\hat V |+_r \tilde-_r 0\rangle = \\ =\langle-_l \tilde-_l|(\hat \sigma_1^z+\hat\sigma_2^z)|+_r \tilde-_r\rangle\langle 1|a^\dagger|0\rangle = - t
\end{multline}
Analogously,
\begin{equation}
 \langle-_l \tilde+_l 0|\hat V |-_r \tilde-_r 1\rangle = \langle-_l \tilde-_l 1|\hat V |-_r \tilde+_r 0\rangle = - t^*.
\end{equation}

\subsubsection{The quadratic corrections arising from non-direct matrix elements.}

Here, we consider the last term in Eq.~\eqref{eff-ham-proj-final}.

We denote
\begin{multline}
 \Delta \hat H_{(2)} = -\frac{1}{2}\sum_{q\in Q}\left(\frac{1}{E_q^{0}-E_p^{0}}+\frac{1}{E_q^{0}-E_{p^\prime}^{0}}\right)\times\\ \times|R_p \rangle\langle L_p|\hat V|R_q\rangle \langle L_q|\hat V|R_{p^\prime}\rangle\langle L_{p^\prime}|,
\end{multline}
\begin{multline}
 \langle L_p|\Delta \hat H_{(2)}|R_{p^\prime}\rangle = 
 -\frac{1}{2}\sum_{q\in Q}\left(\frac{1}{E_q^{0}-E_p^{0}}+\frac{1}{E_q^{0}-E_{p^\prime}^{0}}\right)\times \\ \times\langle L_p|\hat V|R_q\rangle \langle L_q|\hat V|R_{p^\prime}\rangle
\end{multline}

To compute the matrix elements of $\Delta \hat H_{(2)}$, we need to find the possible intermediate $q$-states for the given $p,p^\prime$ matrix element. Below we list the intermediate states in a table.

\begin{table}[h]
\begin{center}
\caption{Allowed intermediate states for the indirect transitions. The first column lists the final states, while the first row lists the initial states.}\label{tab:inter-states}
\begin{tabular}{|c|c|c|c|}\hline
             & $+\tilde-0$ & $-\tilde+0$ & $-\tilde-1$ \\ \hline
 $+\tilde-0$ & $+\tilde-1$, $+\tilde+1$ & $+\tilde+1$ & \\ \hline
 $-\tilde+0$ & $+\tilde+1$ & $-\tilde+1$, $+\tilde+1$ & \\ \hline
 $-\tilde-1$ & & & $-\tilde-0$, $-\tilde-2$, $+\tilde-2$, $-\tilde+2$ \\ \hline
\end{tabular}
\end{center}
\end{table}

The matrix elements read
\begin{multline}
 \langle +_l\tilde-_l0|\Delta \hat H_{(2)}|+_r\tilde-_r0\rangle = -\frac{(s-s^*)^2}{\omega_0} - \frac{(t^*)^2}{\omega_0 + 2\lambda^*} =\\= \frac{4(\im s)^2}{\omega_0}- \frac{(t^*)^2}{\omega_0 + 2\lambda^*},
\end{multline}
\begin{multline}
 \langle -_l\tilde+_l0|\Delta \hat H_{(2)}|-_r\tilde+_r0\rangle = -\frac{(s-s^*)^2}{\omega_0} - \frac{t^2}{\omega_0+2\lambda}=\\=\frac{4(\im s)^2}{\omega_0} - \frac{t^2}{\omega_0+2\lambda},
\end{multline}
\begin{multline}
 \langle +_l\tilde-_l0|\Delta \hat H_{(2)}|-_r\tilde+_r0\rangle = \langle -_l\tilde+_l0|\Delta \hat H_{(2)}|+_r\tilde-_r0\rangle = \\ =
 -\frac{|t|^2}{2}\left(\frac{1}{\omega_0 + 2\lambda^*}+\frac{1}{\omega_0+2\lambda}\right) = \\ = -|t|^2\re\left(\frac{1}{\omega_0+2\lambda}\right).
\end{multline}
\begin{multline}
 \langle -_l\tilde-_l1|\Delta \hat H_{(2)}|-_r\tilde-_r1\rangle =  \\ =\frac{(s+s^*)^2}{\omega_0} -\frac{2(s+s^*)^2}{\omega_0} - \frac{2t^2}{\omega_0+2\lambda} - \frac{2(t^*)^2}{\omega_0+2\lambda^*} =\\= -\frac{4(\re s)^2}{\omega_0} - 4\re\left(\frac{t^2}{\omega_0+2\lambda}\right)
\end{multline}

Here, we took into account that $|\langle 1|a^\dagger|0\rangle|^2=1$ and $|\langle 2|a^\dagger|1\rangle|^2=2$.

\subsubsection{Final form.}

Substituting the results of this section and the expressions for the energies at $g=0$~\eqref{g0-energies} into Eq.~\eqref{eff-ham-proj-final}, we finally obtain the effective Hamiltonian in its matrix form:
\begin{widetext}
\begin{multline}
\langle L_p|\hat H_\mathrm{eff}| R_{p^\prime}\rangle=\\
 \begin{pmatrix}
  2i\im\lambda - g^2\left[\frac{(t^*)^2}{\omega_0+2\lambda^*} - \frac{4(\im s)^2}{\omega_0}\right] & -g^2|t|^2\re\left(\frac{1}{\omega_0+2\lambda}\right) & -gt \\
  -g^2|t|^2 \re\left(\frac{1}{\omega_0+2\lambda}\right) & -2i\im\lambda - g^2\left[\frac{t^2}{\omega_0+2\lambda} - \frac{4(\im s)^2}{\omega_0}\right] & -gt^*\\
-gt & -gt^* & \omega_0 - 2\re \lambda - 4g^2\left[\frac{(\re s)^2}{\omega_0} + \re\left(\frac{t^2}{\omega_0+2\lambda}\right)\right]
 \end{pmatrix}
\end{multline}
\end{widetext}

\section{Solutions of the characteristic polynomial equation.\label{section:ch-pol-sol}}

The three solutions of Eq.~\eqref{ch-pol-depressed} are written explicitly as~\cite{Kurosh-1972}
\begin{align}
\tilde E_1 & = \alpha +\beta,\nonumber\\
\tilde E_2 & = e^{2i\pi/3}\alpha+e^{-2i\pi/3}\beta,\label{eigvals}\\
\tilde E_3 & = e^{-2i\pi/3}\alpha+e^{2i\pi/3}\beta,\nonumber
\end{align}
where
\begin{equation}
\alpha = \sqrt[3]{-q + \sqrt{p^3+q^2}},\label{alphadef}
\end{equation}
\begin{equation}
\beta = -p/\alpha = \sqrt[3]{-q-\sqrt{p^3+q^2}}\label{betadef}
\end{equation}
In the last expression, it is implied that the root branches are chosen in such a way, that the equalities are satisfied.
Finally, the eigenenergies of the effective Hamiltonian matrix can be obtained by transformation~\eqref{shift}.

The combination $(p^3+q^2)$, appearing inside the square roots in Eqs.~\eqref{alphadef} and~\eqref{betadef} is proportional to the discriminant of the depressed cubic~\eqref{ch-pol-depressed} and its sign can be used to characterize the roots of~\eqref{ch-pol-depressed} and, as a consequence, the eigenvalues of the effective Hamiltonian matrix~\cite{Kurosh-1972,Hatsugai-21}:
\begin{itemize}
\item if $(p^3+q^2)>0$, there is one real root and a pair of complex conjugated roots;
\item if $(p^3+q^2)<0$, there are three real roots;
\item if $(p^3+q^2)=0$, there is a double root, which corresponds to a second-order \textit{EP}.
\end{itemize}

The Equation~\eqref{ch-pol-depressed} has a single triple-degenerate root if both coefficients of the depressed cubic are zero: $p=q=0$.
Graphically, we can identify the location of the triple-degeneracy point by looking for the intersection of $p=0$ and $q=0$ lines (see Fig.~\ref{pq-lines}$(a)$). Equivalently, the same point corresponds to the touching of $p^3+q^2=0$ and $p=0$ lines or to the fold in $p^3+q^2=0$ line.

The triple-degeneracy of the eigenvalues is a necessary condition to have a third-order \textit{EP}, but, in principle, it is not sufficient one: the eigenvectors also need to be triple-degenerate. Let $g^\prime/\Omega$ and $\gamma^\prime/\Omega$ be the parameters corresponding to the triple-degenerate eigenvalue $E^\prime$ of the effective Hamiltonian matrix~\eqref{eff-matrix-full}.
The point $(g^\prime\Omega, \gamma^\prime/\Omega)$ will correspond to a third-order \textit{EP}, if the equation on eigenvectors
\begin{equation}
    (H_{p,p^\prime} - E^\prime\delta_{p,p^\prime})u_{p^\prime} = 0
\end{equation}
has only a single solution. As we know from the theory of systems of linear equations, it happens if and only if the matrix $(H_{p,p^\prime} - E^\prime\delta_{p,p^\prime})$ has rank $2$, which is equivalent to the requirement that there is at least one non-zero second-order minor of the matrix~\cite{Kurosh-1972}.

Using the approximated form of the matrix~\eqref{eff-matrix-appr} that works very well in the vicinity of the third-order \textit{EP}, we can compute the minor
\begin{equation}
    \begin{vmatrix}
        H_{1,1}-E & H_{1,2}\\
        H_{3,1} & H_{3,2}
    \end{vmatrix} = -\sqrt{2} i g^\prime\gamma^\prime\sin{2\theta}\neq 0.
\end{equation}
This directly confirms that the triple-degeneracy point corresponding to the intersection of $p=0$ and $q=0$ lines in Fig.~\eqref{pq-lines}$(a)$ is truly a third-order \textit{EP}.

\section{Scaling of the critical values of the interaction strength and the gain/loss parameter.\label{section:critvals}}

In the following, it is convenient to introduce the dimensionless parameters of the interaction strength $\tilde g = g/\Omega$ and gain/loss $\tilde \gamma = \gamma/\Omega$ as well as dimensionless detuning $\Delta=(\omega_r-\Omega)/\Omega$. With the help of these definitions, effective Hamiltonian~\eqref{eff-matrix-appr} reads
\begin{multline}
    H_{p,p^\prime} =\Omega\times\\
    \begin{pmatrix}
        -\frac{2\sin^2{\theta}}{2+\Delta}\times \tilde g^2 & 2i\tilde\gamma\cos{\theta}&-\sqrt{2}\tilde g\sin{\theta}\\
        2i\tilde\gamma\cos{\theta} & 0 & 0 \\
        -\sqrt{2}\tilde g\sin{\theta} & 0 & \Delta - 4\left[\frac{\cos^2{\theta}}{1+\Delta} + \frac{\sin^2{\theta}}{2+\Delta}\right]\times\tilde g^2
    \end{pmatrix}.\label{effham-appr-normalized}
\end{multline}
To make further derivations tractable, we also denote
\begin{equation}
    \frac{2\sin^2{\theta}}{2+\Delta} = u_1,\qquad 4\left[\frac{\cos^2{\theta}}{1+\Delta} + \frac{\sin^2{\theta}}{2+\Delta}\right] = u_2.
\end{equation}
At $\theta,\Delta\ll 1$, $u_1\approx 2\theta^2$ and $u_2\approx 4$.

The critical values of $\tilde g$ and $\tilde \gamma$ correspond to the solution of the system of equations $p,q=0$. We can substitute Eq.~\eqref{cdef} into Eq.~\eqref{ddef} to find
\begin{multline}
    q = \frac{2b^3 - 9cb + 27 d}{54\Omega^3} = \frac{9b(b^2/3 - c) + (27d-b^3)}{54\Omega^3} =\\ = \frac{-27\Omega^2bp + (27d-b^3)}{54\Omega^3}.
\end{multline}
As such, we can conveniently rewrite the system of equations $p,q=0$ as
\begin{align}
    \frac{b^2}{3}-c &= 0,\label{se1_raw}\\
    27d &= b^3.\label{se2_raw}
\end{align}
Directly substituting Eq.~\eqref{effham-appr-normalized} into Eqs.~\eqref{bdef}, \eqref{cdef} and~\eqref{ddef} and then the obtained results into Eqs.~\eqref{se1_raw} and~\eqref{se2_raw}, we arrive at the system
\begin{align}
8\tilde\gamma^2\cos^2\theta = &\frac{2\Delta^2}{3} - \tilde g^2\left[\frac{2\Delta}{3}(2u_2-u_1)-4\sin^2{\theta}\right] +\nonumber\\&+ \tilde g^4\frac{2(u_1^2+u_2^2-u_1u_2)}{3},\label{se1}\\
108\tilde \gamma^2\cos^2{\theta} =& \frac{\left[\Delta - \tilde g^2 (u_1+u_2)\right]^3}{(\Delta - \tilde g^2 u_2)}.\label{se2}
\end{align}

Since at $\theta\ll1$, $u_1\ll u_2$, the right-hand-side of Eq.~\eqref{se2} is very narrowly peaked at $\tilde g^2 = \Delta/u_2$. As such, we can approximate with a good accuracy
\begin{equation}
\tilde g^2_\mathrm{cr} = \frac{\Delta}{u_2}\overset{\Delta,\theta\ll1}{\approx} \frac{\Delta}{4}.
\end{equation}

Substituting it into Eq.~\eqref{se1}, we find then
\begin{equation}
    8 \tilde \gamma_cr^2 \cos^2{\theta} = \frac{4\Delta\sin^2{\theta}}{u_2} + \frac{2\Delta^2}{3}\left(\frac{u_1}{u_2}\right)^2.
\end{equation}
At $\Delta,\theta\ll1$,
\begin{equation}
    \tilde \gamma_\mathrm{cr}^2\approx \frac{\Delta}{8}\times \theta^2.
\end{equation}

Finally, we can go back to dimensional quantities to find
\begin{align}
    g_\mathrm{cr} = \Omega\times \tilde g_\mathrm{cr} &= \frac{\sqrt{\Delta\omega\times \Omega}}{2},\\
    \gamma_\mathrm{cr} = \Omega\times \tilde \gamma_\mathrm{cr} &= \sqrt{\frac{\Delta\omega\times\Omega}{8}}\times \theta = g_\mathrm{cr}\times\frac{\theta}{\sqrt{2}}.
\end{align}

\section{Perturbation of eigenvalues away from the third-order \textit{EP}.\label{section:eigvals-perturb}}

\subsection{Perturbation along the line $p^3+q^2=0$.}

The discriminant equation $p^3+q^2=0$ defines the corresponding curve implicitly. We can turn it into explicit equation
\begin{equation}
    p(q) = -\sqrt[3]{q^2}.\label{discurve}
\end{equation}
The characteristic fold at $p=q=0$ can then be attributed to the singular behaviour of the first derivative $-2/(3\sqrt[3]{q})$ at $q=0$.

When we move away from the point $p=q=0$ along the line~\eqref{discurve}, we observe one singly-degenerate real root
\begin{equation}
 \tilde E_1 = -2\sqrt[3]{q},
\end{equation}
and one doubly-degenerate real root
\begin{equation}
    \tilde E_{2,3} = -2\cos\frac{2\pi}{3}\times \sqrt[3]{q}
\end{equation}

\subsection{Perturbation along the line $q=0$, $p<0$.}

In this case, the parameter $\alpha$ is
\begin{equation}
    \alpha = \sqrt[3]{\sqrt{-|p|^3}} = e^{i\pi/6}\sqrt{|p|}
\end{equation}
And the parameter $\beta$ is
\begin{equation}
\beta = -p/\alpha = e^{-i\pi/6}\sqrt{|p|}
\end{equation}

As such, we observe three distinct real roots
\begin{equation}
 \tilde E_k = 2\cos\left(\frac{\pi}{6}+\frac{2\pi}{3}(k-1)\right)\times\sqrt{|p|},\qquad k=1,2,3.
\end{equation}

\subsection{Perturbation along the line $q=0$, $p>0$.}

In this case,
\begin{align}
\alpha & = \sqrt{p},\\
\beta & = -\sqrt{p}.
\end{align}

We observe one real root $E_1=0$ and a pair of complex conjugated purely imaginary roots
\begin{equation}
E_{2,3} = \pm 2i\sin{\frac{2\pi}{3}}\times\sqrt{p}
\end{equation}

\subsection{Perturbation along the line $q/p=const$.}

For $|q|,|p|\ll1$, we can neglect $p^3$ in comparison with $q^2$. As such, we can write approximately
\begin{equation}
\alpha,\beta = \sqrt[3]{-q\pm|q|}.
\end{equation}
There are three distinct roots then. One real:
\begin{equation}
\tilde E_1 = -\sqrt[3]{2q},
\end{equation}
and a pair of complex conjugated ones
\begin{equation}
E_2 = - e^{2i\pi/3}\sqrt[3]{2q},\qquad E_3 = -e^{-2i\pi/3}\sqrt[3]{2q}.
\end{equation}

\bibliography{sising,circuitQED}

\end{document}